\documentclass[review]{elsarticle}

\usepackage{lineno,hyperref}
\usepackage{amsmath,graphicx, amssymb,tabularx}
\usepackage{subcaption}
\usepackage[font=small,skip=2pt]{caption}
\usepackage{boxedminipage}
\usepackage{algpseudocode}
\usepackage{algorithm}
\usepackage{xcolor}
\modulolinenumbers[5]
\def\x{{\mathbf x}}

\journal{arXiv}

\usepackage{caption}
\begin{document}

\begin{frontmatter}

\title{Tomographic reconstruction to detect evolving structures}

\author[mymainaddress,mysecondaryaddress,mythirdaddress]{Preeti Gopal}

\author[mymainaddress]{Sharat Chandran}

\author[mysecondaryaddress]{Imants Svalbe}
\author[mymainaddress]{Ajit Rajwade}

\address[mymainaddress]{Department of Computer Science and Engineering, IIT Bombay}
\address[mysecondaryaddress]{School of Physics and Astronomy, Monash University}
\address[mythirdaddress]{IITB-Monash Research Academy}
\address{\{preetig,sharat,ajitvr\}@cse.iitb.ac.in,imants.svalbe@monash.edu}

\begin{abstract}
The need for tomographic reconstruction from sparse measurements
arises when the measurement process is potentially harmful, needs to
be rapid, or is uneconomical. In such cases, information from
previous longitudinal scans of the same object (`templates' forming the `prior'), helps to reconstruct the current object (`test') while requiring significantly fewer updating measurements. In this work, we improve the state of the art by proposing the context under which priors can be effectively used based on the final goal of the application at hand.

Our work is based on longitudinal data acquisition scenarios where we wish to study new changes that evolve within an object over time, such as in repeated scanning for disease monitoring, or in tomography-guided surgical procedures. While this is easily feasible when measurements are acquired from a large number of projection angles (referred to as `views' henceforth), it is challenging when the number of views is limited. If the goal is to track the changes while simultaneously reducing sub-sampling artefacts, we propose (1) acquiring measurements from a \emph{small} number of views and using a global unweighted prior-based reconstruction. If the goal is to observe \emph{details} of new changes, we propose (2) acquiring measurements from a \emph{moderate} number of views and using a more involved reconstruction routine. We show that in the latter case, a weighted technique is necessary in order to prevent the prior from adversely affecting the reconstruction of new structures that are absent in any of the earlier scans. The reconstruction of new regions is safeguarded from the bias of the prior by computing regional weights that moderate the local influence of the priors. We are thus able to effectively reconstruct both the old and the new structures in the test. In addition to testing on simulated data, we have validated the efficacy of our method on real tomographic data. The results demonstrate the use of both unweighted and weighted priors in different scenarios. Our methods significantly improve the overall quality of the reconstructed data while minimizing the number of measurements needed for imaging in longitudinal studies.
\end{abstract}


\end{frontmatter}

\section{Introduction}
\label{sec:intro}
Computed Tomography (CT) deals with the recovery of an entire object from a limited set of projection data which are acquired by passing X-rays at different orientations (`views') through the object. In order to minimize the radiation exposed to the subject, current research seeks to significantly reduce the number of measurements required to reconstruct with adequate fidelity. In this regard, there are two lines of pursuit. One is to intelligently choose those sets of projection views that capture most information~\cite{barkan17,fischer16,andrei14}, and the other is to improve the reconstruction algorithms to get the most accurate recovery of the underlying slice, given the measurements from any limited set of views~\cite{yang2018,geyer2015,kilic2011}. This paper deals with the latter scenario.

Measurements in earlier data acquisition techniques are acquired by sampling the physical object greater than the Nyquist rate. If the data is under-sampled i.e., at lesser than Nyquist rate, interpolation methods such as those in~\cite{Shih1992} can be applied. However, these interpolation methods vary with imaging-geometry.

\begin{figure}[!h]
\centering
\fcolorbox{red}{white}{\begin{subfigure}[b]{0.155\linewidth}
        \includegraphics[width=1\textwidth]{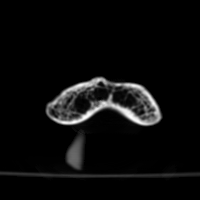}
        \caption{}
     \end{subfigure}
\begin{subfigure}[b]{0.155\linewidth}
        \includegraphics[width=1\textwidth]{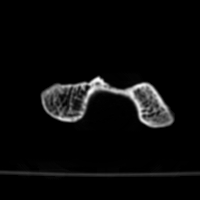}
        \caption{}
     \end{subfigure}
\begin{subfigure}[b]{0.155\linewidth}
        \includegraphics[width=1\textwidth]{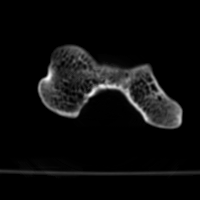}
        \caption{}
     \end{subfigure}
\begin{subfigure}[b]{0.155\linewidth}
        \includegraphics[width=1\textwidth]{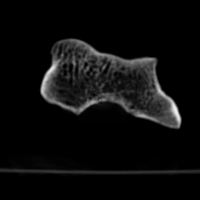}
        \caption{}
     \end{subfigure}
\begin{subfigure}[b]{0.155\linewidth}
        \includegraphics[width=1\textwidth]{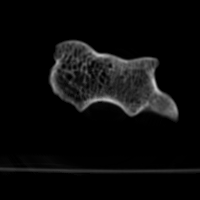}
        \caption{}
     \end{subfigure}
\begin{subfigure}[b]{0.155\linewidth}
        \includegraphics[width=1\textwidth]{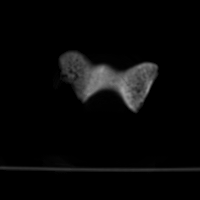}
        \caption{}
     \end{subfigure}}\\
\begin{subfigure}[b]{0.235\linewidth}
        \fcolorbox{yellow}{yellow}{\includegraphics[width=\textwidth]{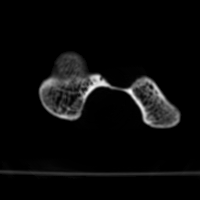}}
        \caption{Test}
\end{subfigure}
\quad
    \begin{subfigure}[b]{0.23\linewidth}
        \includegraphics[width=\textwidth]{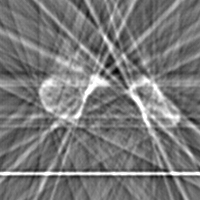}
        \caption{FBP}
     \end{subfigure}
    \begin{subfigure}[b]{0.23\linewidth}
        \includegraphics[width=\textwidth]{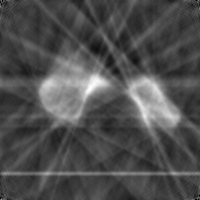}
        \caption{CS}
        \label{fig:CS_blur}
    \end{subfigure}
    \begin{subfigure}[b]{0.23\linewidth}
        \includegraphics[width=\textwidth]{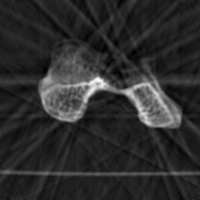}
        \caption{This paper}
        \label{fig:global_cs}
     \end{subfigure}
    \caption{\textit{Knowing more about the data helps.} Reconstructions (h,i,j) of (g) a test slice  of size $(200,200)$ from the Humerus CT dataset~\cite{humerus} are performed using templates [(a)-(f)] and from measurements obtained from only 10 views. (h) Filtered backprojection (FBP) shows streaky sub-sampling artefacts (i) CS reduces the artefacts significantly while slightly blurring details (j) Prior information coupled with CS greatly improves the reconstruction.}
\label{fig:diff_methods}
\end{figure}

In the last decade, a generic (i.e., agnostic to projection-geometry) reconstruction from sub-Nyquist sampled data has been made possible due to methods such as the widely used Compressed Sensing (CS) technique~\cite{Donoho,introCS}, which assumes the data to exhibit properties such as sparsity of the underlying image under certain mathematical transforms such as the Discrete Cosine Transform (DCT) or wavelet transforms. The algorithm presented in~\cite{Varun2013} draws inspiration from the Total Variation (TV) method in CS, and solves for the reconstruction from limited views (upto $20\%$) in micro-CT data using TV with augmented Lagrangian method. Fig.~\ref{fig:CS_blur} shows that although CS removes the artefacts created due to sub-sampling, its reconstruction is blurred depending on the amount of under-sampling.

In addition to CS, when some extra information about the object's structure is known~\cite{PICCS,cardiacPICCS,lubner2011,pirple,mota2017}, it is beneficial in further reducing the number of measurements.  These prior-based techniques use information (`prior') from previously scanned data (`templates') of the same object or a similar one, and utilize it to reconstruct new volumes from a small number of additional measurements.  In techniques such as in~\cite{Ali2018}, reconstruction is performed by imposing wavelet and gradient sparsity in the data, along with the use of Prior Image Constrained Compressed Sensing (PICCS). However, all the above methods have the following two limitations: (1) the prior information may potentially overwhelm the essential details that appear in the test, (2) there is the key issue of choice of \textit{a} particular template among the many previously acquired templates. A large part of this paper is dedicated to alleviating the former limitation, which is an important issue that has so far been overlooked in the literature on tomography. The latter limitation was relaxed in~\cite{liu2016,Xu2012} by building dictionary-based priors from multiple templates. However, as reported in~\cite{my_dicta_paper}, dictionary priors are not as accurate and fast as global eigenspace priors.  Global eigenspace priors are better able to exploit the similarity of a test volume to a set of templates, by assuming that the new test volume lies within the space spanned by the eigenvectors of the multiple representative templates. Fig.~\ref{fig:global_cs} shows the advantage of combining the global prior with CS.

Further, in most of the literature, the reconstructions are shown from projections \textit{simulated} from 3D volumes because the tomographic data are generally proprietary on commercial CT scanners. In this work, we present new datasets consisting of real tomographic measurements and demonstrate our reconstructions on them.

\subsection{Contributions}
\label{sec:contributions}
\begin{figure}[h]
\centering
	\includegraphics[width=0.9\textwidth]{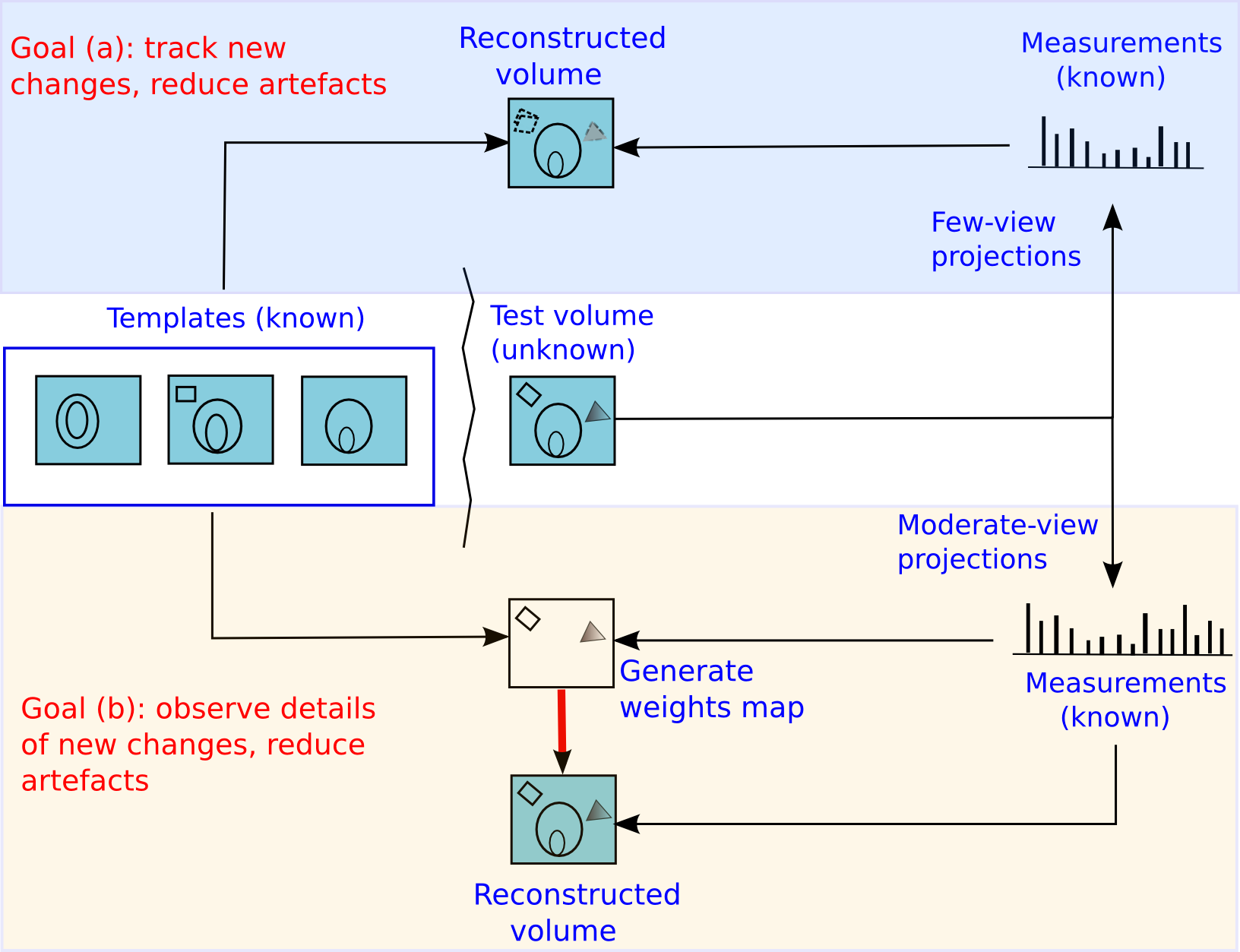}
        \caption{Overview of this paper. The choice of the number of measurement views and the type of reconstruction is driven by the goal in the application under consideration.  When our goal is relatively simple, such as tracking the location of new changes while simultaneously reducing sub-sampling artefacts, we propose (a) acquiring measurements from a small number of views (`few-view' imaging) and using unweighted prior-based reconstruction. When our goal becomes more ambitious, such as observing details of the new changes while simultaneously reducing sub-sampling artefacts, we propose (b) acquiring measurements from a slightly higher number of views (`moderate-view' imaging) and using weighted prior-based reconstruction. In either case, the number of views is lower than what is conventionally used.}
 \label{fig:prior_overview}
\end{figure}

In this work (Fig.~\ref{fig:prior_overview}), we focus on further reducing the number of measurements, with  particular emphasis on longitudinal studies. For example, we consider medical datasets which consist of multiple CT scans being taken during a radio-frequency ablation procedure~\cite{Dong2015}. The process consists of inserting a needle probe into a patient's body. When the needle reaches the tumor site, an electric current is passed to burn the tumor. Repeated CT scans of the patient need to be acquired in order to visualize the movement of the needle and to ensure that it is reaching the appropriate target. A few initial densely-sampled scans are used as templates to help the physician know the position of the needle. The later scans are used to reveal the exact changes during and after burning of the tumor (ablation).  In this context, we demonstrate the combined use of CS with the global prior, in two flavors: the vanilla (unweighted) and weighted global prior-based reconstruction. The choice of the number of measurement views and the type of reconstruction--- unweighted or weighted--- is driven by the goal of the procedure. 

\begin{enumerate}
\item Initially, when our goal is to only track the location of the probe while simultaneously reducing sub-sampling artefacts, we advocate capturing measurements from a small number of views (`few-view' imaging) and using unweighted prior-based reconstruction. 
\item Later, when the probe is proximal to or touches the tumor, our goal is to observe details of tumor ablation while simultaneously reducing sub-sampling artefacts. In this case, we propose to capture measurements from a slightly higher number of views (`moderate-view' imaging) to acquire more information, and use a weighted prior-based reconstruction. The weighted technique moderates the effect of the prior in the reconstruction of new changes in the object being scanned.
\end{enumerate}
  
After the results of the above study are presented, the remainder of this paper discusses each of the above two approaches in detail. Specifically, we discuss how the global prior can impose an inflexible constant weight (and hence a bias) when reconstructing the data. If we want to observe details, this bias can be removed by moderating the control of the prior by imposing spatially varying weights to the prior. In addition to the ablation data, we have also validated our method on real tomographic datasets of longitudinal studies on biological specimen.

This paper is organized as follows: Section~\ref{sec:tmh} demonstrates the utility of both the unweighted and weighted methods on a longitudinal medical dataset. Section~\ref{sec:unweighted_global_prior} describes the construction of the unweighted global eigenspace prior. In addition, we recap the advantages of global prior over dictionary priors. Section~\ref{sec:weighted_prior} describes how the unweighted prior needs to be modified when accurate details of new changes are to be observed. A weighted technique offers a solution. Results validated on both synthetic and real 3D biological datasets are shown. Finally, we conclude with key inferences that can be drawn from our work in Section~\ref{sec:conclusions}.


\section{Application: Reconstruction for CT-guided radio-ablation study}
\label{sec:tmh}
Before diving into the details of the unweighted and the weighted global prior methods, we first show how both the techniques can be applied to our advantage in a real-life medical longitudinal study. Our data~\footnote{Source: Tata Memorial Centre~\cite{tmh}, Parel, Mumbai.  This is the national comprehensive centre for the prevention, treatment, education and research in cancer, and is recognized as one of the leading cancer centres in India.} consists of CT scans from a longitudinal study. This involves successive scans of the liver taken during a radio-frequency ablation procedure described in the previous section. In such a procedure, the physician inserts a thin needle-like probe into the organ~\cite{rfa}. Once the needle hits the tumor, a high-frequency current is passed through the tip of the probe and this burns the malignant tumor. Throughout this process, multiple CT scans help the physician to track the position of the needle and check the changes within. In this context, we classify the goal of any of our reconstruction techniques into two categories: 

\begin{figure}[h!]
    \begin{subfigure}[b]{0.24\linewidth}
        \includegraphics[width=\textwidth]{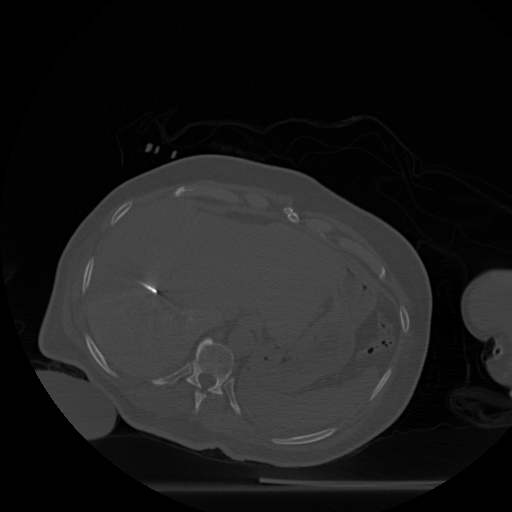}
 \caption{slice 1}
    \end{subfigure}
    \begin{subfigure}[b]{0.24\linewidth}
        \includegraphics[width=\textwidth]{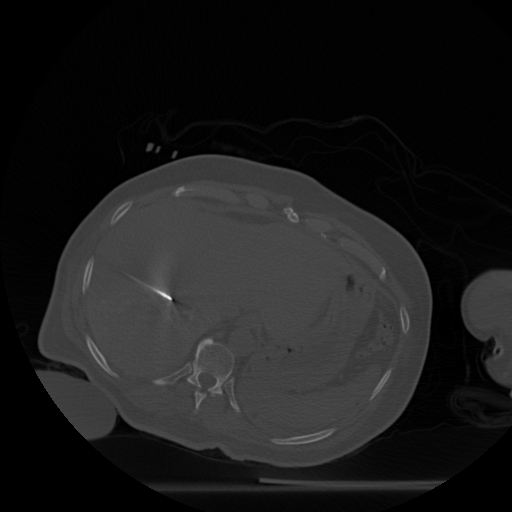}
 \caption{slice 2}
    \end{subfigure}
     \begin{subfigure}[b]{0.24\linewidth}
        \includegraphics[width=\textwidth]{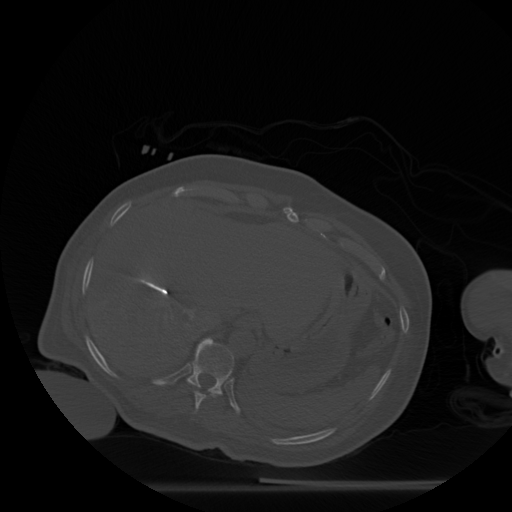}
 \caption{slice 3}
    \end{subfigure}
       \begin{subfigure}[b]{0.24\linewidth}
        \includegraphics[width=\textwidth]{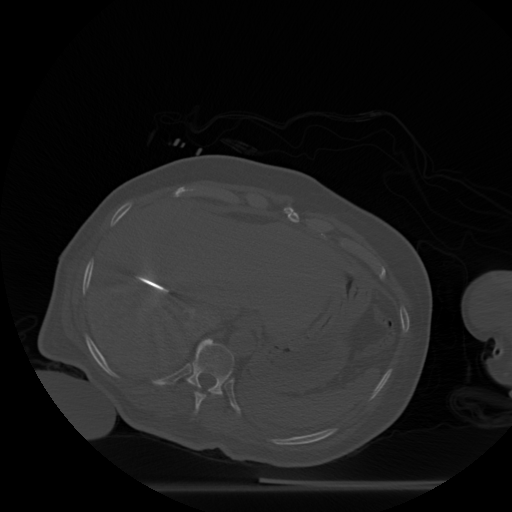}
 \caption{slice 4}
    \end{subfigure}
       \begin{subfigure}[b]{0.24\linewidth}
        \includegraphics[width=\textwidth]{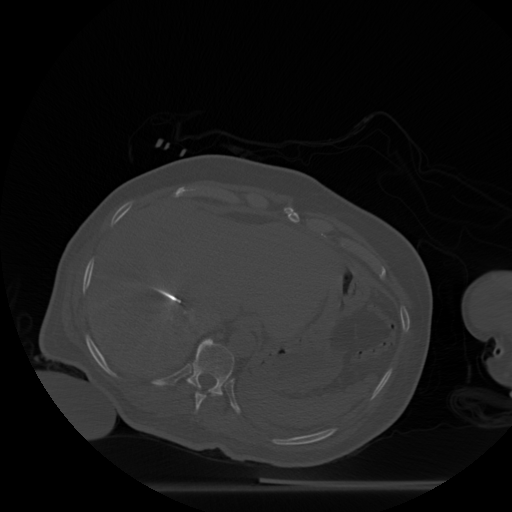}
 \caption{slice 5}
    \end{subfigure}
              \begin{subfigure}[b]{0.24\linewidth}
        \includegraphics[width=\textwidth]{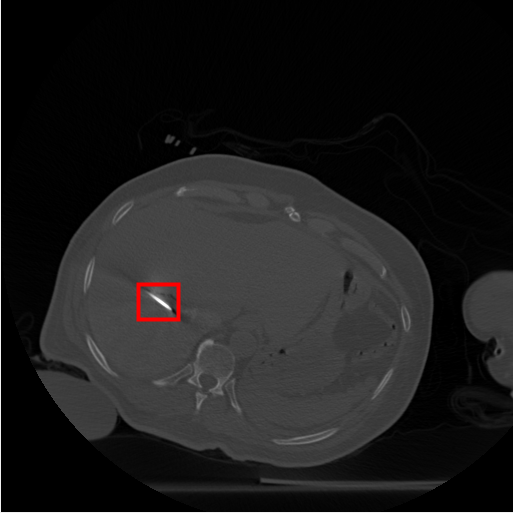}
 \caption{slice 6}
    \end{subfigure}
       \begin{subfigure}[b]{0.24\linewidth}
        \includegraphics[width=\textwidth]{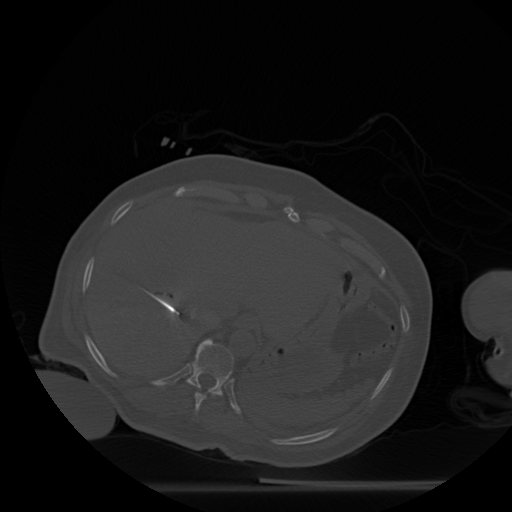}
 \caption{slice 7}
    \end{subfigure}
       \begin{subfigure}[b]{0.24\linewidth}
        \includegraphics[width=\textwidth]{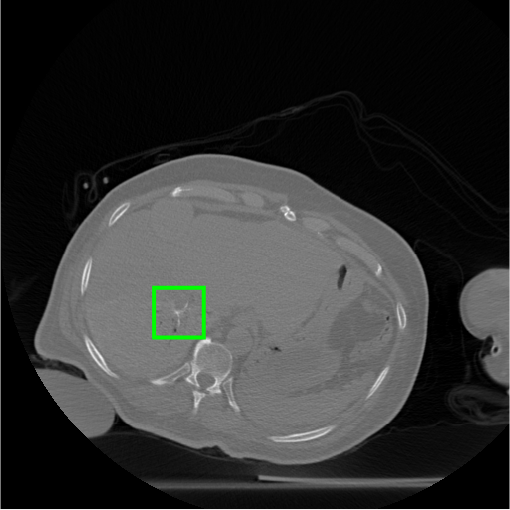}
 \caption{slice 8}
    \end{subfigure}     
     \caption{Radio-frequency ablation dataset. (a) through (g): one of the slices (512 x 512) from each of the 8 scan volumes of a longitudinal study dataset of the liver. Note that in volumes (a) through (g), the needle (shown in red in (f)) approaches the target tumor. (h) the organ after the ablation: this slice is displayed on a separate intensity scale to enable proper viewing of the region marked in green that shows the after-effects of ablating the tumor.} 
\label{fig:RFA2_test_templates}
\end{figure}

\begin{figure}[!h]
\centering
\subcaptionbox{Test}{\fcolorbox{white}{green}{\includegraphics[width=0.23\columnwidth]{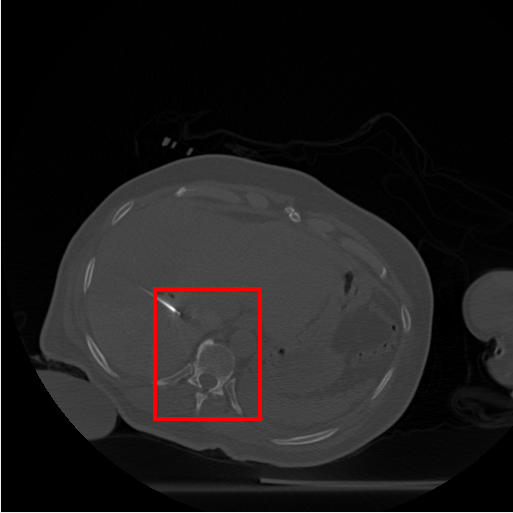}}}\hfill
\subcaptionbox{Backprojection}{\includegraphics[width=0.24\columnwidth]{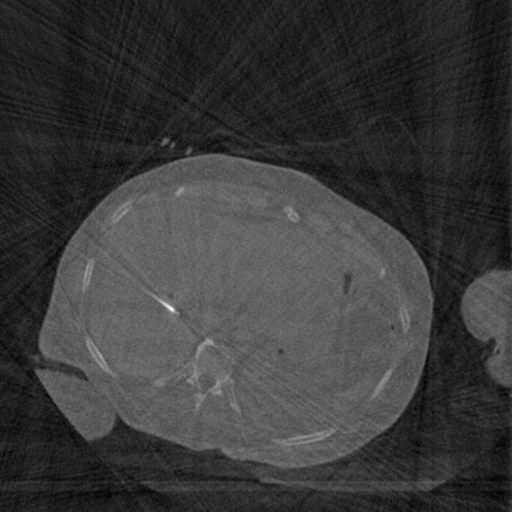}}\hfill
\subcaptionbox{CS}{\includegraphics[width=0.24\columnwidth]{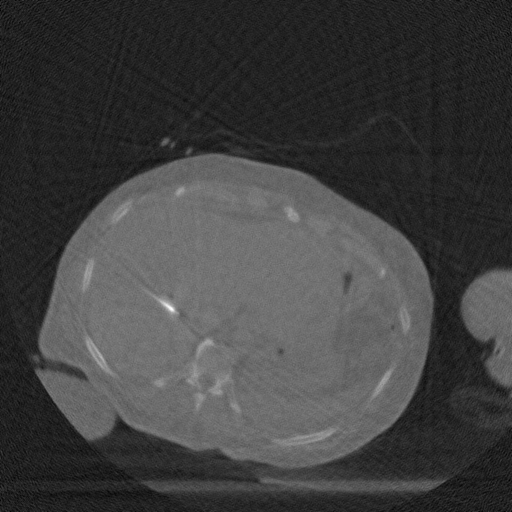}}\hfill
\subcaptionbox{Proposed\\ method: unweighted prior}{\includegraphics[width=0.24\columnwidth]{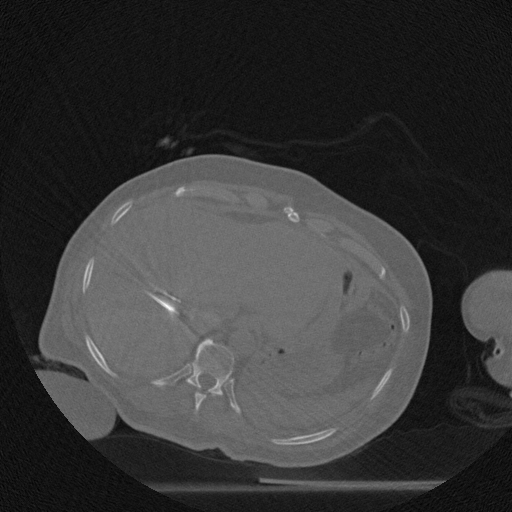}}
\caption[Representative results-1]{\textbf{Goal: Track new changes.}\small{ Reconstruction of slice 7 (`test') of Fig.~\ref{fig:RFA2_test_templates} from only 90 views, using (b) Filtered backprojection and no prior resulting in streaks, the Structural Similarity Index Metric (SSIM) $= 0.48$ (c) CS and no prior resulting in blurred bone structures, SSIM $=0.35$ and (d) unweighted global prior (slices 1-6 of Fig.~\ref{fig:RFA2_test_templates} are used as templates) resulting in clear bone structures with less streaks, SSIM $=0.55$. The region enclosed in red rectangle is our RoI as it contains both the new position of the needle and some background. All SSIM values are computed for this RoI.}}
\label{fig:RFA2_very_few_views}
\end{figure}


\begin{figure}[!h]
\centering
\subcaptionbox{Test}{\fcolorbox{white}{blue}{\includegraphics[width=0.23\columnwidth]{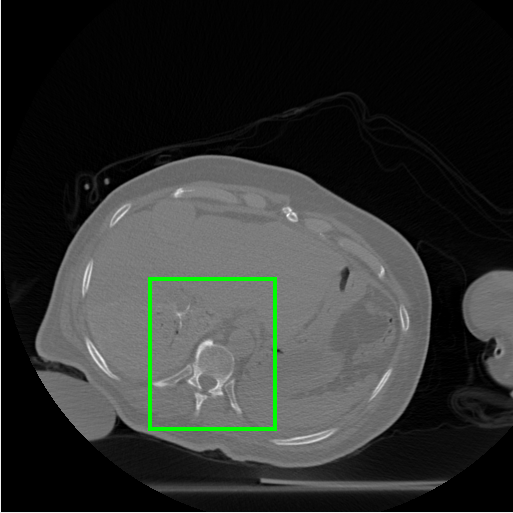}}}\hfill
\subcaptionbox{Backprojection}{\includegraphics[width=0.24\columnwidth]{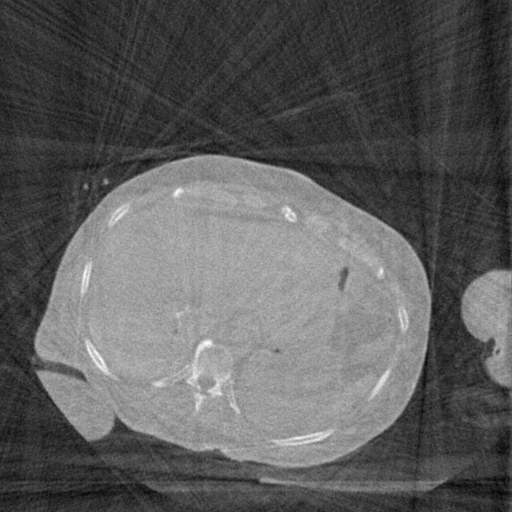}}\hfill
\subcaptionbox{CS}{\includegraphics[width=0.24\columnwidth]{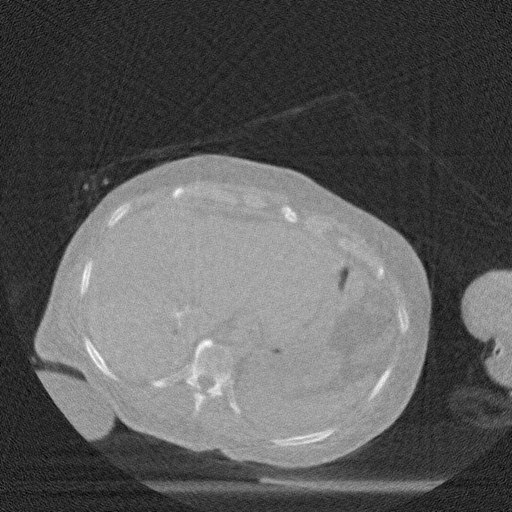}}\hfill
\subcaptionbox{Proposed\\ method: unweighted\\ prior}{\includegraphics[width=0.24\columnwidth]{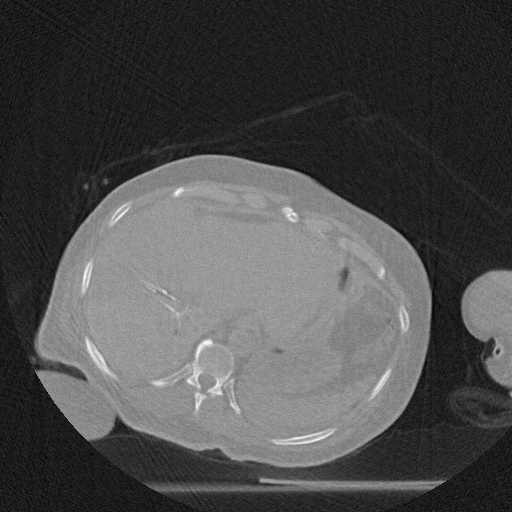}}
\subcaptionbox{Proposed\\ method: weighted prior}{\includegraphics[width=0.24\columnwidth]{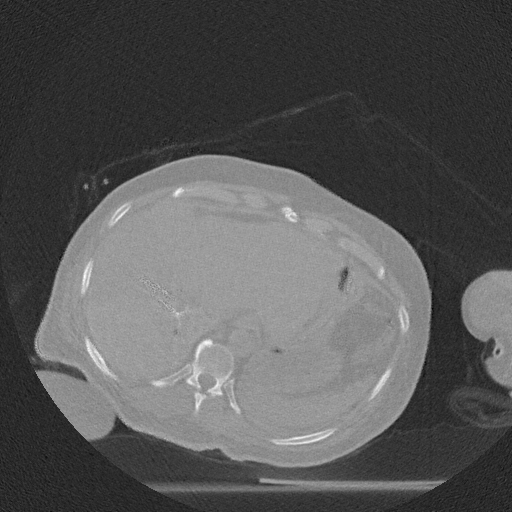}}
\subcaptionbox{Weights map}{\includegraphics[width=0.24\columnwidth]{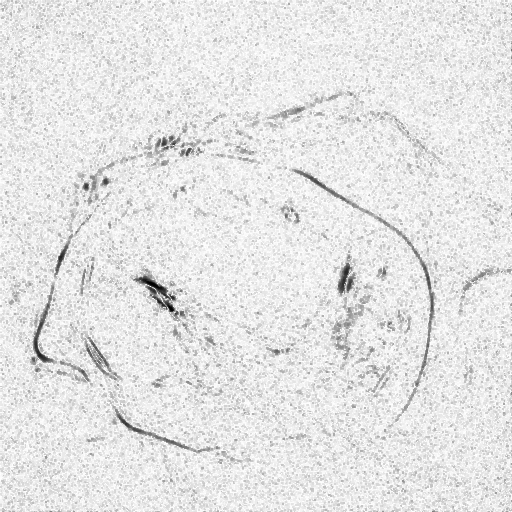}}
\caption[Representative results-2]{\textbf{Goal: Track new changes.} Reconstruction of slice 8 of Fig.~\ref{fig:RFA2_test_templates} from 120 views, using (b) FBP with SSIM $= 0.50$ (c) CS with SSIM $=0.46$ and (d) unweighted global prior, with SSIM $=0.51$ (\emph{notice dominance of the prior: a prominent residual shadow of the needle which was present in the prior templates, but not present in the test image}), and (e) weighted prior with SSIM $=0.56$ (\emph{notice that the dominance of the prior is significantly controlled}). The region enclosed in green rectangle is our Region of Interest (RoI) as it contains both the new position of the needle and some background. The SSIM is computed in this RoI. (f) shows the computed weights map (described later in the paper) used for reconstruction. Darker intensities indicate lower weights to prior as these are the regions of new changes.}
\label{fig:RFA2_few_views}
\end{figure}

\begin{enumerate}
\item To track the position of the needle in a relatively well reconstructed background.
\item To accurately observe the new changes amidst a relatively well reconstructed background after the needle touches the tumor.
\end{enumerate} 

The needle has a very high attenuation coefficient when compared to that of the organs. Hence, the needle can be tracked by acquiring measurements from a very small number of views. We use the unweighted global prior reconstruction here to reduce the artefacts due to sub-sampling. The unweighted method is fast and sufficient to track the position of the needle.  Once the needle reaches the site of the tumor, we propose changing the imaging protocol to acquire measurements from a moderate number of views. This will enable us to get more information about the new changes. In addition, we then deploy the weighted prior method in order to locate the regions of new changes and penalize any dominance of the prior in these regions.  Regardless of the imaging protocol we use (`few' or `moderate'), the number of views is smaller (upto one-fifth) than the conventional number of views used in a standard hospital setting.

The dataset from this longitudinal medical study consists of 8 scans taken during the ablation procedure. We demonstrate our method for 2D reconstruction by choosing a single slice from each of the 8 volumes as our dataset. Note that all these 8 slices are located at the \emph{same} index~\footnote{The notion of~\textit{same index} (slice number corresponding to the same depth) makes sense in the context, because in such problems, the different scans are aligned with each other.} within each of the respective volumes. Fig.~\ref{fig:RFA2_test_templates} shows the chosen set of 2D slices (each of size $512 \times 512$) from the different volumes. Observe that the needle is seen in all of the first 7 slices  and the effect of ablation is seen in the 8th slice. 

\textbf{Tracking the needle:} We first choose slices 1-6 as our templates, and reconstruct slice 7 with the specific goal of tracking the needle and simultaneously reduce artefacts. Fig.~\ref{fig:RFA2_very_few_views} shows the reconstruction of slice 7 from its measurements from only 90 views. The reconstructions are quantitatively compared using SSIM. 

\textbf{Observing details of the ablation:} Next, we choose slices 1-7 as our templates and reconstruct slice 8 from 120 views i.e., a somewhat higher number of views this time. Fig.~\ref{fig:RFA2_few_views} shows the reconstructions of slice 8 by different methods. We see that the weighted prior reconstruction brings in the advantage of the prior without it adversely affecting the new regions.

\section{Methods}
\label{sec:methods}
Having presented the application, we first review the algorithm~\cite{my_dicta_paper} for a global (unweighted) prior-based reconstruction in Sec.~\ref{sec:unweighted_global_prior}. This is followed by details of our technique in Sec.~\ref{sec:weighted_prior}. Our method estimates the location and magnitude of new changes, and eventually prevents the prior from adversely affecting the reconstruction of new regions in the test. This, we refer to as a weighted global prior-based reconstruction routine.
\subsection{\textbf{Summary of unweighted global prior-based reconstruction}}
\label{sec:unweighted_global_prior}

Principal Component Analysis (PCA) has been traditionally used to find the significant modes of (Gaussian corrupted) data. In this regard, it has been widely applied in the context of data compression. However, PCA can also be seen as a tool to provide an orthogonal basis to represent the space in which the test data could lie. This space is constructed from the available set of templates which must cover a realistically representative range of structures. We first present the eigenspace-cum-CS prior-based reconstruction, which was shown~\cite{my_dicta_paper} to be better when compared to dictionary-based priors. 

To begin with, when an object is scanned multiple times, a set of high quality reconstructions (i.e., reconstructions from a dense set of projection views) may be chosen as templates for the reconstruction of future scan volumes, which in turn, may be scanned using far fewer measurements. The eigenspace $E_{\text{high}}$ of the $L$ prior
templates $Q_1,Q_2,...,Q_L$ is pre-computed. Here, it is assumed that the test volume can be expressed as a compact linear combination of the principal components extracted from a set of similar volumes. Hence, the prior is built using PCA. For the eigenspace to encompass a range of possible structures in the test slice, the templates must represent a wide structural range. Moreover, if these volumes are not aligned, then they must be first registered before computing the prior. The prior is built by computing the covariance matrix from the template set $\{Q_i\}_{i=1}^L$. \emph{The space spanned by the eigenvectors $\{\boldsymbol{V_k}\}_{k=1}^{L-1}$ (eigenspace) of the covariance matrix is the global prior and is assumed to contain any test slice that is similar, but not necessarily identical to the templates.}  We use all of the $\textrm{L}-1$ orthogonal eigenvectors as a basis to represent the unknown test volume. Let $\boldsymbol{x} = \boldsymbol{\Psi\theta}$ denote the reconstructed volume, $\boldsymbol{y} = \boldsymbol{\Phi x}$ its measured tomographic projections, $\boldsymbol{\Phi}$ the tomographic projection operator, $\boldsymbol{\theta}$ the sparse coefficients of $\boldsymbol{x}$, $\boldsymbol{\Psi}$ the basis in which $\boldsymbol{x}$ is assumed to be sparse, $\boldsymbol{\mu}$ the mean of the templates, and $\boldsymbol{\alpha}$ the vector of eigen-coefficients of the test scan. Then, once the eigenspace is pre-computed, the test is reconstructed by minimizing the following cost function:
 \begin{equation}
   \setlength{\belowdisplayskip}{0pt} \setlength{\belowdisplayshortskip}{0pt}
\setlength{\abovedisplayskip}{-2pt} \setlength{\abovedisplayshortskip}{-2pt}
J_1(\boldsymbol{\theta,\alpha}) = \lVert\boldsymbol{\Phi x}- \boldsymbol{y}\rVert_2^2  + \lambda_1\lVert\boldsymbol{\theta}\rVert_1+\lambda_2\lVert\boldsymbol{x} - (\boldsymbol{\mu} + \sum_{k}\boldsymbol{V_k}\alpha_k)\rVert_2^2.
\label{Eq:main_prior}
\end{equation}
Here, $\lambda_1$, $\lambda_2$ are tunable weights given to the sparsity and prior terms respectively. The unknowns $\boldsymbol{\theta}$ and $\boldsymbol{\alpha}$
are solved by alternately minimizing $J_{\boldsymbol{\alpha}}(\boldsymbol{\theta})$ using a fixed $\boldsymbol{\alpha}$, and $J_{\boldsymbol\theta}(\boldsymbol{\alpha})$ using the resultant $\boldsymbol{\theta}$, where 
\begin{align}
J_{\boldsymbol{\alpha}}(\boldsymbol{\theta}) &\triangleq \lVert\boldsymbol{\Phi x- y}\rVert_2^2  + \lambda_1\lVert\boldsymbol{\theta}\rVert_1+\lambda_2\lVert\boldsymbol{x} - (\boldsymbol{\mu + V\alpha})\rVert_2^2, \\
J_{\boldsymbol\theta}(\boldsymbol{\alpha}) &\triangleq \lVert\boldsymbol{\Psi\theta} - (\boldsymbol{\mu + V\alpha})\rVert_2^2.
\end{align}
Note that $\boldsymbol{\theta}$ is solved for using the basis pursuit CS solver~\cite{l1ls}. Solving for $\boldsymbol{\alpha}$ leads to the closed form update:
\begin{equation}
\boldsymbol{\boldsymbol{\alpha}} = \boldsymbol{V}^T(\boldsymbol{\Psi \theta} -\boldsymbol{\mu}).
\end{equation}
 Optimal values of $\lambda_1, \lambda_2$ must be empirically chosen \textit{a priori}, based on the reconstructions of one of the template volumes. The cost function described in Eq.~\ref{Eq:main_prior} is biconvex and the convergence of this optimization is guaranteed by the monotone convergence theorem~\cite{monotone_convergence_theorem}.

\subsection{\textbf{Weighted prior-based reconstruction}}
\label{sec:weighted_prior}
Although the unweighted global prior can be very useful in some circumstances, as was shown in Sec.~\ref{sec:tmh}, it poses a major limitation when we want accurate details of the new changes. While the unweighted prior compensates very well for the possible artefacts due to sparse measurements, it dominates the regions with new changes masking the signal, as seen in Fig.~\ref{fig:RFA2_few_views}-d. Ideally, we will want to impose the prior only in the regions that are common between the test and templates.  Our weighted prior based reconstruction overcomes this limitation by minimizing the following cost function:
\begin{equation}
J_3(\boldsymbol{\theta},\boldsymbol{\alpha}) = \lVert\boldsymbol{\Phi x}-\boldsymbol{y}\rVert_2^2  + \lambda_1\lVert\boldsymbol{\theta}\rVert_1 +\lambda_2\lVert\boldsymbol{W}(\boldsymbol{x} - (\boldsymbol{\mu} + \sum_{k}\boldsymbol{V_k}\alpha_k))\rVert_2^2.
\label{eq:weighted_prior}
\end{equation}
The key to our method is the discovery of a diagonal weights
 matrix $\boldsymbol{W}$, where $W_{ii}$ contains the (non-negative) weight assigned to the $i^{\textrm{th}}$ voxel of the prior. $\boldsymbol{W}$ is first constructed using some preliminary reconstruction methods (to be described in the following section), following which Eq.~\ref{eq:weighted_prior} is used to obtain the final reconstruction. In regions of change in test data, we want lower weights for the prior when compared to regions that are similar to the prior.  

\subsubsection{\textbf{Computation of weights matrix $\boldsymbol{W}$}}
Since the test volume (referred to as $\boldsymbol{x}$) is unknown to begin with, it is not possible to decipher the precise regions in $\boldsymbol{x}$ that are different from all the templates. We refer the reader to Schematic 1 that describes the evolution of the procedure used to detect the new regions in the unknown volume. We start with $X^{\text{fdk}}$, the initial backprojection reconstruction of the test volume using the Feldkamp-Davis-Kress (FDK) algorithm~\cite{FDK} in an attempt to discover the difference between the templates and the test volume. Let $\boldsymbol{V_{\text{high}}}$ be the eigenspace constructed from high-quality templates. However, the difference between $X^{\text{fdk}}$ and its projection onto the eigenspace $\boldsymbol{V_{\text{high}}}$ will detect the new regions along with  many false positives (false new regions). This is because, $X^{\text{fdk}}$ will contain many geometric-specific artefacts arising from sparse measurements (angle undersampling), which are absent in the high quality templates used to construct the eigenspace $\boldsymbol{V_{\text{high}}}$. To discover unwanted artefacts of the imaging geometry, in a counter-intuitive way, we generate \emph{low quality} reconstruction of the templates as described below.

\subsubsection{\textbf{Algorithm to compute weights-map $\boldsymbol{W}$}}
\label{sec:thealgo}
\begin{enumerate}

\item Perform a pilot reconstruction $X^{\text{fdk}}$ of the
  test volume $\boldsymbol{x}$ using FDK.

\item Compute low quality template volumes $Y^\text{fdk}$. 
In Schematic 1, for ease of exposition, we assumed a single
template. In the sequel, we assume $L$ templates from which we build
an eigenspace. 
\vspace{-0.1cm}

  \begin{enumerate}
  \item Generate simulated measurements $\boldsymbol{y_{Q_i}}$ for every template $Q_i$, using the exact same projections views and imaging geometry with which the measurements $\boldsymbol{y}$ of the test volume $\boldsymbol{x}$ were acquired, and 
\item Perform $L$  preliminary FDK reconstructions of each of the $L$ templates from $\boldsymbol{y_{Q_i}}$.  Let this be denoted by $\{Y^{\text{fdk}}_i\}_{i=1}^L$.
  \end{enumerate}
\item Build eigenspace $\boldsymbol{V_{\text{low}}}$ from $\{Y^{\text{fdk}}_i\}_{i=1}^L$.  Let $P^{\text{fdk}}$ denote projection of $X^{\text{fdk}}$ onto $\boldsymbol{V_{\text{low}}}$. The difference between $P^{\text{fdk}}$ and $X^{\text{fdk}}$ will not contain false positives due to imaging geometry, but will have false positives due to artefacts that are specific to the reconstruction method used. To resolve this, perform steps $4$ and $5$.
\item Project with multiple methods.
  \begin{enumerate}
  \item Perform pilot reconstructions of the test using $M$ different
    reconstruction algorithms\footnote{CS~\cite{lasso}, Algebraic Reconstruction Technique (ART)~\cite{art},
     Simultaneous Algebraic Reconstruction Technique (SART)~\cite{sart} and  Simultaneous Iterative Reconstruction Technique (SIRT)~\cite{sirt}}. Let this set be denoted
    as $X \triangleq \{X^j\}_{j=1}^M$ where $j$ is an index for the
    reconstruction method, and $X^1 = X^{\text{fdk}}$. 

  \item From $\boldsymbol{y_{Q_i}}$, perform reconstructions of the template $Q_i$ using the $M$ different algorithms, for each of the $L$ templates. Let this set be denoted by $Y \triangleq \{\{Y_{i}^j\}_{j=1}^M\}_{i=1}^L$ where $Y^{1}_i = Y^{\text{fdk}}_i$, $\forall i \in \{1,..,,L\}$.

  \item For each of the $M$ algorithms (indexed by $j$), build an eigenspace $\boldsymbol{V_\text{low}^j}$ from $\{Y_1^j,Y_2^j, \ldots, Y_{L}^j\}$. 

  \item Next, for each $j$,  project $X^j$  onto $\boldsymbol{V_\text{low}^j}$. Let this projection be denoted by $P^j$. To reiterate, this captures those parts of the test volume that lie in the subspace $\boldsymbol{V_\text{low}^j}$ (i.e., are similar to the template reconstructions). The rest, i.e., new changes and their reconstruction method-dependent-artefacts, are not captured by this projection and need to be eliminated.
  \end{enumerate}
\item To remove all reconstruction method dependent false positives, we compute $\min_{j}(|X^j(x,y,z) - P^j(x,y,z)|)$. (The intuition for using the `min' is provided in the paragraph immediately following step 6 of this procedure.)
\item Finally, the weight to prior for each voxel coordinate $(x,y,z)$ is given by
  \begin{equation} 
    \boldsymbol{W_v}(x,y,z) = (1+k(\min_{j}|X^j(x,y,z) - P^j(x,y,z)|))^{-1}.
  \end{equation}
\label{eq:weights}
\end{enumerate}
\vspace{-0.3in}
Note that here $\boldsymbol{W_v}(x,y,z)$ represents the weight to the prior in the $(x,y,z)^{th}$ voxel. $\boldsymbol{W_v}(x,y,z)$ must be
low whenever the preliminary test reconstruction $X^j(x,y,z)$ is different from its projection $P^j(x,y,z)$ onto the prior eigenspace, for \emph{every} method $j \in \{1,...,M\}$. This is
because it is unlikely (details in Sec.~\ref{sec:motivation_multiple_eigenspaces}) that \emph{every} algorithm would produce a significant artefact at a voxel, and hence we hypothesize that the large difference has arisen due to genuine structural changes. The parameter $k$ decides the sensitivity of the weights to the difference $|X^j(x,y,z) - P^j(x,y,z)|$ and hence it depends on the size of the new regions we want to detect.  
We found that our final reconstruction results obtained by
solving Eqn.~\ref{eq:weighted_prior} were robust over a wide range of $k$ values, as discussed in Sec.~\ref{subsec:k}.

\noindent \begin{boxedminipage}{\textwidth} 
{\bf Schematic 1}: Motivation behind our algorithm. (The plus $\oplus$ and the
minus $\ominus$ operators are placeholders; precise details available
in Section~\ref{sec:thealgo}). 

{\small
 Let \textcolor{blue}{prior $Q:=$ old regions ($O$)}\\
 Let \textcolor{blue}{test volume $\boldsymbol{x}:=$ old regions ($O$) $\oplus$ new regions ($N$)}\\
    \begin{enumerate}
\item  Compute pilot reconstruction of $\boldsymbol{x}$. Let this be called $X$.\\ \textcolor{blue}{$X = O \oplus N \oplus Ar(O) \oplus Ar(N)$}, where \\ \textcolor{blue}{$Ar(O)$} denote the reconstruction artefacts that depend on the old regions, the imaging geometry and the reconstruction method, and\\
 \textcolor{blue}{$Ar(N)$} denote the reconstruction artefacts that depend on the new regions, imaging geometry and the reconstruction method.
\item  Note that \textcolor{blue}{$Q \ominus X = N \oplus Ar(O) \oplus Ar(N)$} gives the new regions, but along with lots of artefacts due to the imaging geometry (sparse views). To eliminate these unwanted artefacts, compute \textcolor{blue}{$Y = Q \oplus Ar(O)$} by simulating projections from $Q$ using the same imaging geometry used to scan $\boldsymbol{x}$, and then reconstructing a lower quality prior volume $Y$. 
    \item Note that \textcolor{blue}{$Y \ominus X = N \oplus Ar(N)$} contains the artefacts due to the new regions only. These are different for different reconstruction methods. To eliminate these method dependent artefacts, compute $Y$ and $X$ using different reconstruction methods. Let these be denoted by $Y^j$ and $X^j$ respectively.
    \item Compute
\vspace{-0.2cm}
           \textcolor{blue}{\begin{equation*}
            Y^1 \ominus X^1 = N \oplus Ar^1(N)
           \end{equation*}
\vspace{-0.5cm}
           \begin{equation*}
            Y^2 \ominus X^2 = N \oplus Ar^2(N)
           \end{equation*}}
\vspace{-0.4cm}
   \item  New regions are obtained by computing
\vspace{-0.2cm}
          \textcolor{blue}{\begin{equation*}
          (Y^1 \ominus X^1)\cap(Y^2 \ominus X^2)= N
          \end{equation*}}   
\vspace{-0.5cm}
    \item Finally, assign space-varying weights $\boldsymbol{W}$ based on step $5$.
    \end{enumerate}
\label{algo:newRegionDetection}
}
\end{boxedminipage}

\subsubsection{Motivation for the use of multiple types of eigenspaces for the computation of weights}
\label{sec:motivation_multiple_eigenspaces}

\begin{figure}[!h]
    \begin{subfigure}[b]{0.24\linewidth}
        \includegraphics[width=\textwidth]{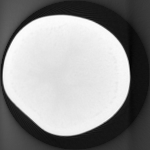}
\captionsetup{labelformat=empty}       
 \caption{}
    \end{subfigure}
    \begin{subfigure}[b]{0.24\linewidth}
        \includegraphics[width=\textwidth]{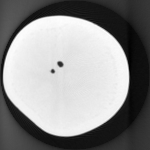}
\captionsetup{labelformat=empty}
        \caption{}
     \end{subfigure}
    \begin{subfigure}[b]{0.24\linewidth}
        \includegraphics[width=\textwidth]{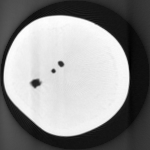}
\captionsetup{labelformat=empty}
        \caption{}
     \end{subfigure}
    \begin{subfigure}[b]{0.235\linewidth}
        \fcolorbox{yellow}{yellow}{\includegraphics[width=\textwidth]{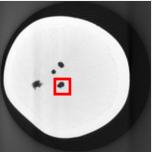}}
\captionsetup{labelformat=empty}
        \caption{}
\label{fig:potato_test}
     \end{subfigure}
      \caption{Potato dataset: One slice (slice-A) each from the templates
        (the first three from left) and a slice from the test 
        volume (extreme right). Notice the appearance of the fourth
        hole in the test slice. }
\label{fig:templates_test_potato_A1}
\end{figure}

\begin{figure}[!h]
    \begin{subfigure}[b]{0.24\linewidth}
        \includegraphics[width=\textwidth]{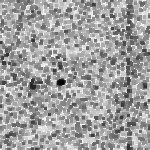}
        \caption{FBP}
    \end{subfigure}
    \begin{subfigure}[b]{0.24\linewidth}
        \includegraphics[width=\textwidth]{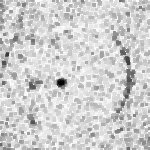}
        \caption{CS-DCT}
     \end{subfigure}
    \begin{subfigure}[b]{0.24\linewidth}
        \includegraphics[width=\textwidth]{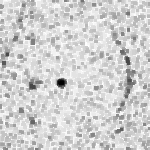}
        \caption{CS-Haar}
     \end{subfigure}
    \begin{subfigure}[b]{0.24\linewidth}
        \includegraphics[width=\textwidth]{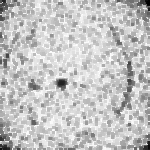}
        \caption{ART}
     \end{subfigure}
    \begin{subfigure}[b]{0.24\linewidth}
        \includegraphics[width=\textwidth]{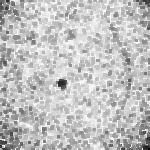}
        \caption{SART}
     \end{subfigure}
    \begin{subfigure}[b]{0.24\linewidth}
        \includegraphics[width=\textwidth]{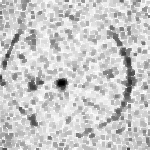}
        \caption{SIRT}
     \end{subfigure}
    \begin{subfigure}[b]{0.24\linewidth}
        \includegraphics[width=\textwidth]{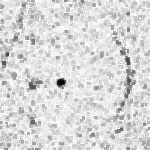}
        \caption{combined}
     \end{subfigure}
      \caption{Weight maps (corresponding to the difference between pilot reconstruction of the image in the last sub-figure of Fig.~\ref{fig:templates_test_potato_A1} and its projection onto the eigenspace $V_{low}$) constructed using different reconstruction methods, as specified in Eq.~\ref{eq:weights}. The weight maps are different because the reconstruction artefacts of the new structures in test image will be different for every reconstruction method used, as seen in Fig.~\ref{fig:reconstructions_diff_methods}. }
\label{fig:weights_map_2Dpotato}
\end{figure}
\begin{figure}[!h]
    \begin{subfigure}[b]{0.24\linewidth}
        \includegraphics[width=\textwidth]{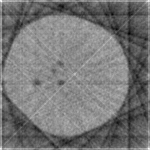}
        \caption{FBP}
    \end{subfigure}
    \begin{subfigure}[b]{0.24\linewidth}
        \includegraphics[width=\textwidth]{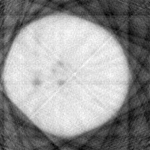}
        \caption{CS-DCT}
     \end{subfigure}
    \begin{subfigure}[b]{0.24\linewidth}
        \includegraphics[width=\textwidth]{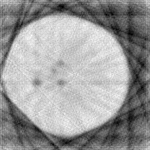}
        \caption{CS-Haar}
     \end{subfigure}
    \begin{subfigure}[b]{0.24\linewidth}
        \includegraphics[width=\textwidth]{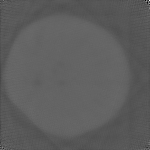}
        \caption{ART}
     \end{subfigure}
    \begin{subfigure}[b]{0.24\linewidth}
        \includegraphics[width=\textwidth]{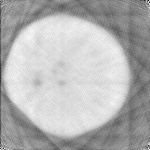}
        \caption{SART}
     \end{subfigure}
    \begin{subfigure}[b]{0.24\linewidth}
        \includegraphics[width=\textwidth]{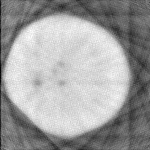}
        \caption{SIRT}
     \end{subfigure}
    \begin{subfigure}[b]{0.24\linewidth}
        \includegraphics[width=\textwidth]{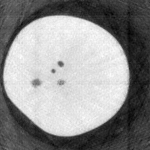}
        \caption{combined}
     \end{subfigure}
      \caption{(a)-(g):Different reconstructions of~\ref{fig:potato_test}. The magnitude and sharpness of the artefacts is different for each method. (h) Weighted-prior method that combines weights-map information from all other methods.} 
\label{fig:reconstructions_diff_methods}
\end{figure}

The changes and new structures present in the test data will generate different artifacts for different reconstruction techniques. These artifacts would not be captured by reconstructions of the templates since the underlying new changes and structures may be absent in all of the templates. We aim to let the weights be independent of the type of artifact. Hence, we use a combination of different reconstruction techniques to generate different types of eigenspaces and combine information from all of them to compute weights. To illustrate the benefit of this method, we first performed 2D reconstruction of a test slice from the potato dataset (please refer Sec.~\ref{Sec:potato} for details of the dataset) Fig.~\ref{fig:templates_test_potato_A1} shows the test and template slices. Fig.~\ref{fig:weights_map_2Dpotato}  shows the weight maps generated using Eq~\ref{eq:weights} by various reconstruction methods. It can be seen that the weights are low in the region of the new change in test data. Because all the iterative methods are computationally expensive, we chose only FBP and CS-DCT for computing weight maps for all 3D reconstructions.

\section{Results and Discussion}

The proposed methods have been validated on 2D and 3D synthetic and real tomographic data of  biological and medical specimens. As with any global prior-based method, there is a need for the test volume to be registered with the templates. In all our longitudinal study experiments, the volumes were already aligned while imaging. However, if this were not the case, then the prior must be aligned to the test based on an initial pilot reconstruction of the test. The results below demonstrate the advantage of the global prior method over the patch-based method, and the advantage of the weighted global prior method over the unweighted method.

\subsection{\textbf{Evaluation of weighted prior-reconstruction in real 3D data}}


The proposed method has been validated on new\footnote {This data and our code will be made available to the community.} scans of biological specimens in a longitudinal setting. These datasets in the form of \emph{raw cone-beam projection measurements} were acquired from a lab at the Australian National University (ANU). We emphasize that in most of the literature on tomographic reconstruction, the results are shown on reconstruction from projections \emph{simulated} from 3D volumes. This is because most CT scanners do not reveal the raw projections, and instead output only the full reconstructed volumes. Moreover, the process of conversion from the projections to the full volumes is proprietary. \emph{Departing from this, we demonstrate reconstruction results from raw projection data.} In all figures in this section, `unweighted prior' refers to optimizing Eqn.~\ref{eq:weighted_prior} with $\boldsymbol{W}(x,y,z)=1$.
\begin{figure}[!h]
    \begin{subfigure}[b]{0.24\linewidth}
        \includegraphics[width=\textwidth]{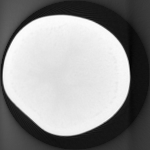}
\captionsetup{labelformat=empty}       
 \caption{}
    \end{subfigure}
    \begin{subfigure}[b]{0.24\linewidth}
        \includegraphics[width=\textwidth]{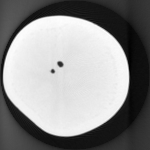}
\captionsetup{labelformat=empty}
        \caption{}
     \end{subfigure}
    \begin{subfigure}[b]{0.24\linewidth}
        \includegraphics[width=\textwidth]{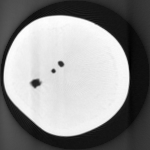}
\captionsetup{labelformat=empty}
        \caption{}
     \end{subfigure}
    \begin{subfigure}[b]{0.235\linewidth}
        \fcolorbox{blue}{blue}{\includegraphics[width=\textwidth]{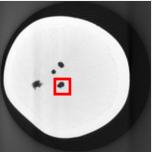}}
\captionsetup{labelformat=empty}
        \caption{}
     \end{subfigure}
      \caption{Potato 3D dataset: One slice (slice-A) each from the templates
        (the first three from left) and a slice from the test 
        volume (extreme right). Notice the appearance of the fourth
        hole in the test slice. }
\label{fig:templates_test_potato_A}
\end{figure}

\begin{figure}[!h]
\centering
\subcaptionbox{Test}{\fcolorbox{blue}{blue}{\includegraphics[width=0.185\columnwidth]{images/potato/testIm_color.png}}}\hfill
\subcaptionbox{FDK, no prior}{\includegraphics[width=0.19\columnwidth]{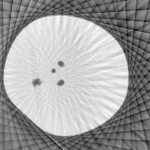}}\hfill
\subcaptionbox{CS, no prior}{\includegraphics[width=0.19\columnwidth]{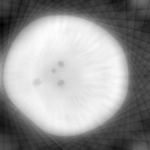}}\hfill
\subcaptionbox{Unweighted \\ prior}{\includegraphics[width=0.19\columnwidth]{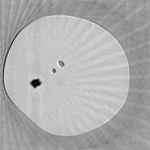}}\hfill
\subcaptionbox{Weighted\\ prior}{\includegraphics[width=0.19\columnwidth]{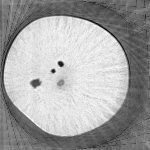}}
\caption{Slice-A from 3D reconstruction of the potato with $5\%$ projection views--(b) has strong streak artefacts with unclear shadow of the potato, (c) largely blurred, (d) no new information detected (prior dominates) and (e) new information detected while simultaneously reducing streak artefacts. The reconstructed volumes can be viewed in~\cite{supp_paper}.}
\label{fig:potato_3D_results_A}
\end{figure}

\subsubsection{\textbf{Potato dataset}}
\label{Sec:potato}
The first (Potato) dataset consisted of four scans of the humble potato, chosen for its simplicity (Figs.~\ref{fig:templates_test_potato_A}). Measurements from each scan consisted of real cone-beam projections from 900 views, each of size $150\times150$. The corresponding size of the reconstructed volume is $150\times150\times100$. While the first scan was taken of the undistorted potato, subsequent scans were taken of the same specimen, each time after drilling a new hole halfway into the potato.  Projections were obtained using circular cone beam geometry. The specimen was kept in the same position throughout the acquisitions. In cases where such an alignment is not present, all the template volumes must be pre-aligned before computing the eigenspace. The test must be registered to the templates after its preliminary pilot reconstruction. 
 The ground truth consists of FDK reconstructions from the full set of acquired measurements from 900 projection views. The test volume was reconstructed using measurements from 45 projection views, i.e, $5\%$ of the projection views from which ground truth was reconstructed.  The selected 3D
ground truth of template volumes, test volume, as well as the 3D reconstructions are shown in~\cite{supp_paper}. Fig.~\ref{fig:potato_3D_results_A} shows a slice from the reconstructed 3D volume. We observe that our method reconstructs new structures while simultaneously reducing streak artefacts.\\ 
\begin{table}[!h]
\caption{SSIM of 3D reconstructed potato volumes from various methods. Each RoI spans 7 consecutive slices where the test is different from all of the templates. In this dataset alone, the new changes are in a homogeneous background. Hence, the FDK performs the best when the RoI alone is considered. However, it fails when the entire volume is considered due to the prominent streaky artefacts.}
\begin{tabular}{|l|c|c|c|c|c|}
\hline
                   & \textbf{\begin{tabular}[c]{@{}c@{}}ground\\truth\end{tabular}} & \textbf{FDK} & \textbf{CS} & \textbf{\begin{tabular}[c]{@{}c@{}}Unweighted\\prior + CS\end{tabular}} & \textbf{\begin{tabular}[c]{@{}c@{}}Weighted\\prior + CS\end{tabular}} \\ \hline
\textbf{red RoI}   & 1 (ideal)                     & \textcolor{red}{0.939}        & 0.850       & 0.712                                                               & 0.852                                                            \\ \hline
\textbf{full volume} & 1 (ideal)                     & 0.744        & 0.817       & 0.856                                                               & \textcolor{red}{0.857}                                                                  \\ \hline
\end{tabular}
\label{table:potato_ssim}
\end{table}
\subsubsection{\textbf{Okra dataset}}
\label{Sec:okra}
\begin{figure}[!h]
    \begin{subfigure}[b]{0.18\linewidth}
        \includegraphics[width=\textwidth]{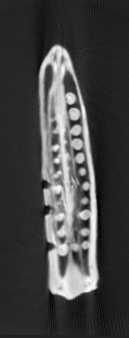}
\captionsetup{labelformat=empty}       
 \caption{}
    \end{subfigure}
    \begin{subfigure}[b]{0.18\linewidth}
        \includegraphics[width=\textwidth]{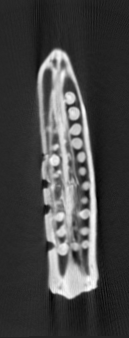}
\captionsetup{labelformat=empty}
        \caption{}
     \end{subfigure}
    \begin{subfigure}[b]{0.18\linewidth}
        \includegraphics[width=\textwidth]{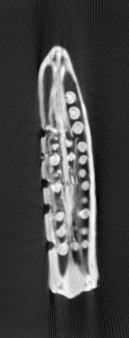}
\captionsetup{labelformat=empty}
        \caption{}
     \end{subfigure}
    \begin{subfigure}[b]{0.18\linewidth}
        \includegraphics[width=\textwidth]{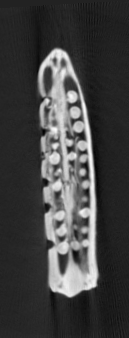}
\captionsetup{labelformat=empty}
        \caption{}
     \end{subfigure}
    \begin{subfigure}[b]{0.176\linewidth}
        \fcolorbox{yellow}{yellow}{\includegraphics[width=\textwidth]{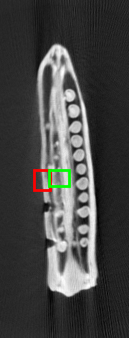}}
\captionsetup{labelformat=empty}
        \caption{}
     \end{subfigure}
     \caption{Okra 3D dataset: One slice each from the templates (the
       first four from the left), and one from the test volume
       (extreme right). In the regions marked in red and green, while
       all slices have deformities, the test
       has none.}
\label{fig:templates_test_okra}
\end{figure}

\begin{figure}[!h]
\centering
\subcaptionbox{Test}{\fcolorbox{yellow}{yellow}{\includegraphics[width=0.185\columnwidth]{images/okra/testCropped.png}}}\hfill
\subcaptionbox{FDK, no prior}{\includegraphics[width=0.19\columnwidth]{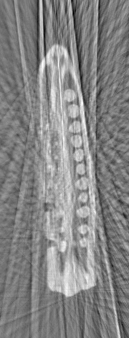}}\hfill
\subcaptionbox{CS, no prior}{\includegraphics[width=0.19\columnwidth]{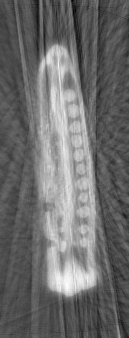}}\hfill
\subcaptionbox{Unweighted \\ prior}{\includegraphics[width=0.19\columnwidth]{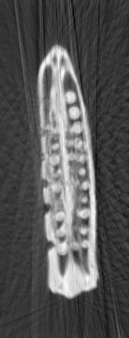}}\hfill
\subcaptionbox{Weighted\\ prior}{\includegraphics[width=0.19\columnwidth]{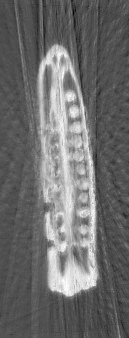}}
\caption{3D reconstruction of the okra from $10\%$ projection
  views (b) has strong streak artefacts, (c) blurred, (d) no new
  information detected (prior dominates -- the deformity from the prior
  shows up as a false positive) and (e) new information detected (no deformities
  corresponding to red and green regions) while simultaneously
  reducing streak artefacts. The reconstructed volumes can be viewed in ~\cite{supp_paper}.}
\label{fig:okra_3D_results}
\end{figure}

In order to test on data with intricate structures, a \textbf{second (Okra) dataset} consisting of five scans of an okra specimen was acquired (Fig.~\ref{fig:templates_test_okra}). The measurements consisted of real cone-beam projections from 450 views, each of size $336\times156$. The corresponding size of the reconstructed volume is $338\times338\times123$. Prior to the first scan, two holes were drilled on the surface of the specimen. This was followed by four scans, each after introducing one new cut. The ground truth consists of FDK reconstructed volumes from the the full set of 450 view projections. The test volume was reconstructed from a partial set of 45
projections, i.e, $10\%$ of the projection views from which ground truth was reconstructed. The selected 3D ground truth of template volumes, the test volume as well as the 3D reconstructions can be seen in~\cite{supp_paper}. One of the slices of the reconstructed volumes is shown in Fig.~\ref{fig:okra_3D_results}. The red and green 3D RoI in the video and images show the regions where new changes are present. Based on the potato and the okra experiments, we see that our method is able to \emph{discover both} the presence of a new structure, as well as the absence of a structure.

\begin{figure}[!h]
\centering
\subcaptionbox{Test}{\includegraphics[width=0.19\columnwidth]{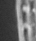}}\hfill
\subcaptionbox{FDK, no prior}{\includegraphics[width=0.19\columnwidth]{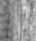}}\hfill
\subcaptionbox{CS, no prior}{\includegraphics[width=0.19\columnwidth]{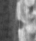}}\hfill
\subcaptionbox{Unweighted \\ prior}{\includegraphics[width=0.19\columnwidth]{images/okra/zoomed/pca_cropped.png}}\hfill
\subcaptionbox{Weighted\\ prior}{\includegraphics[width=0.19\columnwidth]{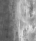}}
\caption{Zoomed in portion corresponding to the red RoI of Fig.~\ref{fig:okra_3D_results} for various methods (b) has strong streak artefacts, (c) blurred, (d) no new  information detected (prior dominates -- the deformity from the prior
  shows up as a false positive) and (e) new information detected (no deformities
 ).}
\label{fig:okra_zoomed_3D_results}
\end{figure}

\begin{table}[!h]
\caption{SSIM of 3D RoI of reconstructed okra volumes from various methods. Each RoI spans 7 consecutive slices where the test is different from all of the templates.}
\begin{tabular}{|l|c|c|c|c|c|}
\hline
                   &  \textbf{\begin{tabular}[c]{@{}c@{}}Ground\\truth\end{tabular}} & \textbf{FDK} & \textbf{CS} & \textbf{\begin{tabular}[c]{@{}c@{}}Unweighted\\prior + CS\end{tabular}} & \textbf{\begin{tabular}[c]{@{}c@{}}Weighted\\prior + CS\end{tabular}} \\ \hline
\textbf{red RoI}   & 1 (ideal)                     & 0.737        & 0.836       & 0.858                                                               & \textcolor{red}{0.883}                                                                  \\ \hline
\textbf{green RoI} & 1 (ideal)                     & 0.798        & \textcolor{red}{0.861}       & 0.800                                                               & \textcolor{red}{0.861}                                                                  \\ \hline
\end{tabular}
\label{table:okra_ssim}
\end{table}


\begin{figure}[!h]
    \begin{subfigure}[b]{0.3\linewidth}
        \includegraphics[width=\textwidth]{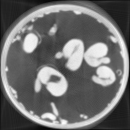}
\captionsetup{labelformat=empty}
        \caption{}
    \end{subfigure}
\quad
    \begin{subfigure}[b]{0.3\linewidth}
        \includegraphics[width=\textwidth]{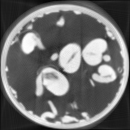}
\captionsetup{labelformat=empty}
        \caption{}
     \end{subfigure}
\quad
    \begin{subfigure}[b]{0.3\linewidth}
        \includegraphics[width=\textwidth]{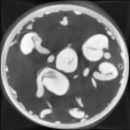}
\captionsetup{labelformat=empty}
        \caption{}
     \end{subfigure}\\
    \begin{subfigure}[b]{0.3\linewidth}
        \includegraphics[width=\textwidth]{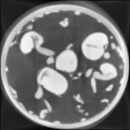}
\captionsetup{labelformat=empty}
        \caption{}
     \end{subfigure}
\quad
    \begin{subfigure}[b]{0.3\linewidth}
        \includegraphics[width=\textwidth]{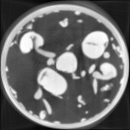}
\captionsetup{labelformat=empty}
        \caption{}
     \end{subfigure}
\quad
    \begin{subfigure}[b]{0.29\linewidth}
        \fcolorbox{green}{green}{\includegraphics[width=\textwidth]{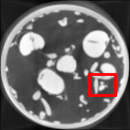}}
\captionsetup{labelformat=empty}
        \caption{}
     \end{subfigure}
      \caption{Sprouts 3D dataset: One slice each from the templates
        (the first five from left) and a slice from the test (extreme
        right).}
\label{fig:templates_test_sprouts}
\addtolength{\textfloatsep}{-0.8cm}
\end{figure}
\subsubsection{\textbf{Sprouts dataset}}
\label{Sec:sprouts}
The third dataset consists of six scans of an in vivo
sprout specimen imaged at its various stages of growth
(Fig.~\ref{fig:templates_test_sprouts}). Projections were generated from the given volume of size $130\times130\times130$. In contrast to the scientific
experiment performed for the case of the okra and the potato where we
introduced man-made defects, the changes here are purely the work of
nature. 
\begin{figure}[!h]
    \begin{subfigure}[b]{0.29\linewidth}
        \fcolorbox{green}{green}{\includegraphics[width=\textwidth]{images/sprouts/testIm_red.png}}
\captionsetup{labelformat=empty}
        \caption{Test}
     \end{subfigure}
\quad
    \begin{subfigure}[b]{0.3\linewidth}
        \includegraphics[width=\textwidth]{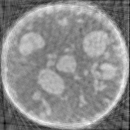}
\captionsetup{labelformat=empty}
        \caption{FDK}
    \end{subfigure}
\quad
    \begin{subfigure}[b]{0.3\linewidth}
        \includegraphics[width=\textwidth]{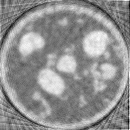}
\captionsetup{labelformat=empty}
        \caption{CS}
     \end{subfigure}\\
\quad
    \begin{subfigure}[b]{0.3\linewidth}
        \includegraphics[width=\textwidth]{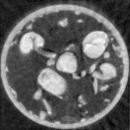}
\captionsetup{labelformat=empty}
        \caption{Unweighted Prior}
     \end{subfigure}
\quad
    \begin{subfigure}[b]{0.3\linewidth}
        \includegraphics[width=\textwidth]{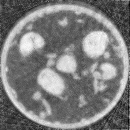}
\captionsetup{labelformat=empty}
        \caption{Weighted prior}
     \end{subfigure}
     \caption{3D reconstruction of sprouts from $2.5\%$ projection views
   (b, c) have poor details (d) no new information detected (the
   prior dominates as can be seen in the blue and red regions) and
   (e) new information detected in the regions of interest. The reconstructed volumes can be viewed in ~\cite{supp_paper}.} 
\label{fig:sprouts_3D_results}
\end{figure}
\begin{table}[]
\caption{SSIM of 3D RoI of reconstructed sprouts volumes from various methods. The RoI spans 7 consecutive slices where the test is different from all of the templates.}
\begin{tabular}{|l|c|c|c|c|c|}
\hline
                   & \textbf{\begin{tabular}[c]{@{}c@{}}ground\\truth\end{tabular}} & \textbf{FDK} & \textbf{CS} & \textbf{\begin{tabular}[c]{@{}c@{}}Unweighted\\prior + CS\end{tabular}} & \textbf{\begin{tabular}[c]{@{}c@{}}Weighted\\prior + CS\end{tabular}} \\ \hline
\textbf{red RoI}   & 1 (ideal)                     & 0.852        & 0.843       & 0.834                                                               & \textcolor{red}{0.881}                                                                  \\ \hline
\end{tabular}
\label{table:sprouts_ssim}
\end{table}
The ground truth consists of FDK reconstructed volumes from a set of
1800 view projections. The test volume was reconstructed from partial
set of 45 projections, i.e, $2.5\%$ of the projection views from which
ground truth was reconstructed. The selected 3D ground truth of
template volumes, test volume, as well as the 3D reconstructions are
shown in~\cite{supp_paper}. One of the slices of the
reconstructed volumes is shown in
Fig.~\ref{fig:sprouts_3D_results}. As an example, the red region of interest (RoI) has been culled out from 7
consecutive slices in the 3D volume to indicate new structures; other
changes can be viewed in the video. Tables~\ref{table:potato_ssim},~\ref{table:okra_ssim} and \ref{table:sprouts_ssim} shows the improvement in the Structure
Similarity Index (SSIM) of the reconstructed new regions as compared
to other methods. 

\subsection{\textbf{More results on CT-guided radio-frequency ablation data}}

\begin{figure}[!h]
       \begin{subfigure}[b]{0.24\linewidth}
        \includegraphics[width=\textwidth]{images/tmh/RFA2/template1.png}
\captionsetup{labelformat=empty}       
 \caption{image-1}
    \end{subfigure}
       \begin{subfigure}[b]{0.24\linewidth}
        \includegraphics[width=\textwidth]{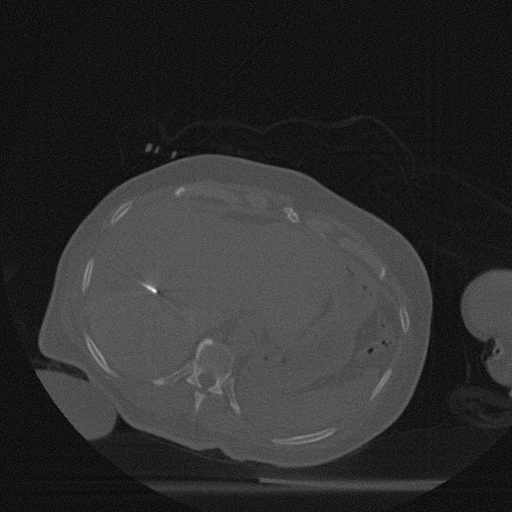}
\captionsetup{labelformat=empty}       
 \caption{image-1 recon}
    \end{subfigure}
 \begin{subfigure}[b]{0.24\linewidth}
        \includegraphics[width=\textwidth]{images/tmh/RFA2/template2.png}
\captionsetup{labelformat=empty}       
 \caption{image-2}
    \end{subfigure}
       \begin{subfigure}[b]{0.24\linewidth}
        \includegraphics[width=\textwidth]{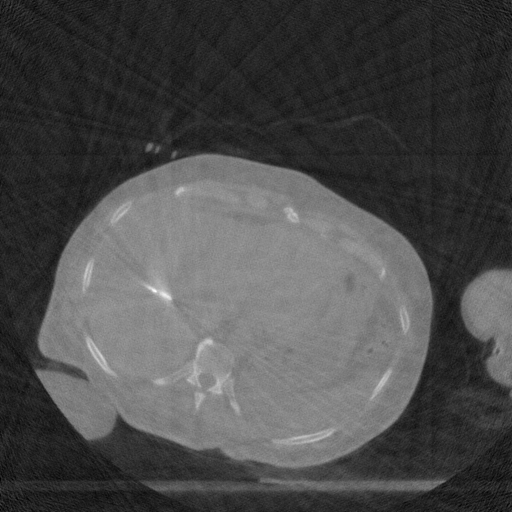}
\captionsetup{labelformat=empty}       
 \caption{image-2 recon}
    \end{subfigure}
 \begin{subfigure}[b]{0.24\linewidth}
        \includegraphics[width=\textwidth]{images/tmh/RFA2/template3.png}
\captionsetup{labelformat=empty}       
 \caption{image-3}
    \end{subfigure}
       \begin{subfigure}[b]{0.24\linewidth}
        \includegraphics[width=\textwidth]{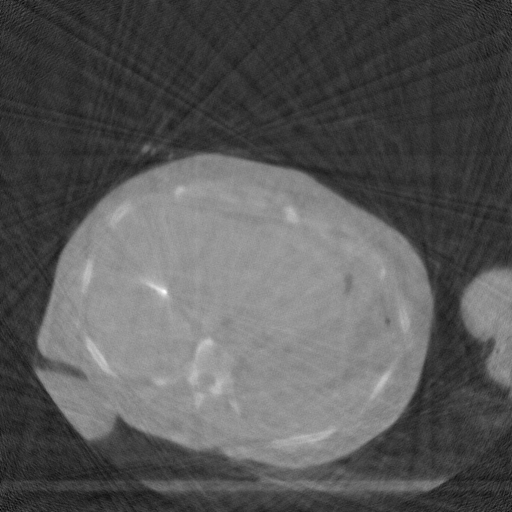}
\captionsetup{labelformat=empty}       
 \caption{image-3 recon}
    \end{subfigure}
 \begin{subfigure}[b]{0.24\linewidth}
        \includegraphics[width=\textwidth]{images/tmh/RFA2/template4.png}
\captionsetup{labelformat=empty}       
 \caption{image-4}
    \end{subfigure}
       \begin{subfigure}[b]{0.24\linewidth}
        \includegraphics[width=\textwidth]{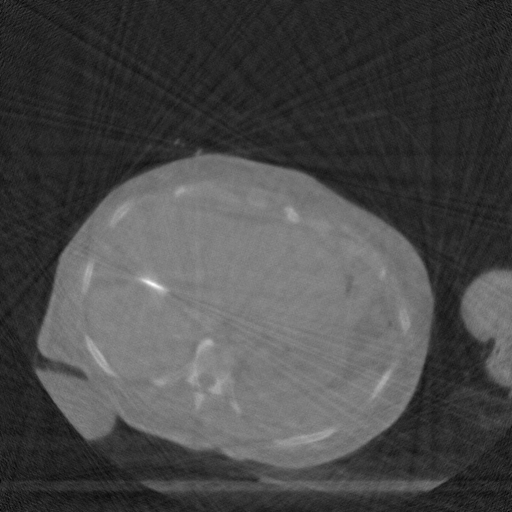}
\captionsetup{labelformat=empty}       
 \caption{image-4 recon}
    \end{subfigure}
 \begin{subfigure}[b]{0.24\linewidth}
        \includegraphics[width=\textwidth]{images/tmh/RFA2/template5.png}
\captionsetup{labelformat=empty}       
 \caption{image-5}
    \end{subfigure}
       \begin{subfigure}[b]{0.24\linewidth}
        \includegraphics[width=\textwidth]{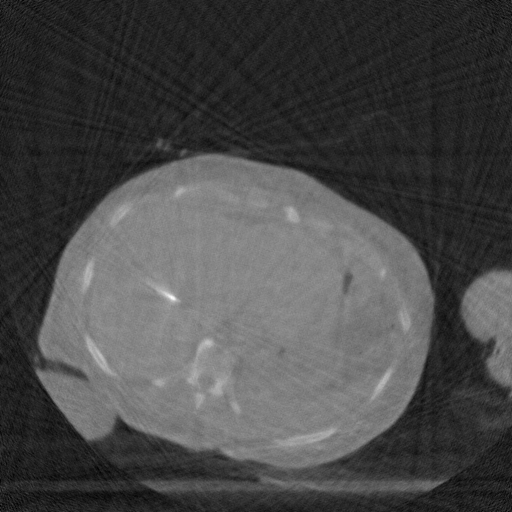}
\captionsetup{labelformat=empty}       
 \caption{image-5 recon}
    \end{subfigure}
 \begin{subfigure}[b]{0.24\linewidth}
        \includegraphics[width=\textwidth]{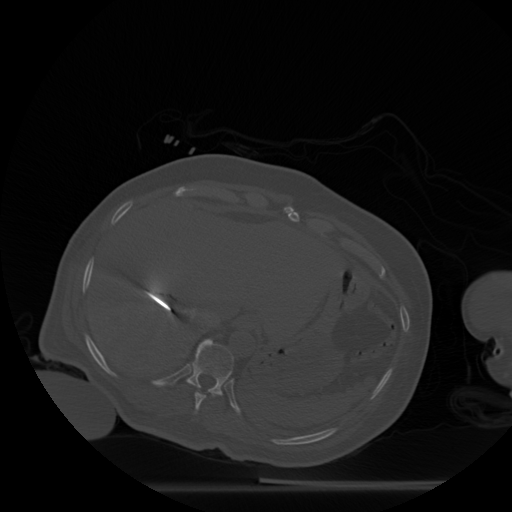}
\captionsetup{labelformat=empty}       
 \caption{image-6}
    \end{subfigure}
       \begin{subfigure}[b]{0.24\linewidth}
        \includegraphics[width=\textwidth]{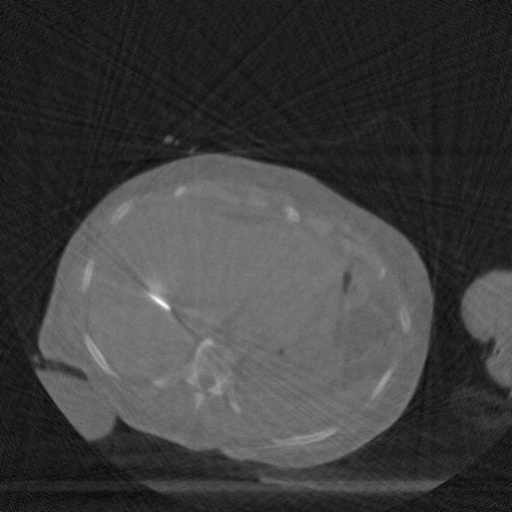}
\captionsetup{labelformat=empty}       
 \caption{image-6 recon}
    \end{subfigure}
 \begin{subfigure}[b]{0.24\linewidth}
        \includegraphics[width=\textwidth]{images/tmh/RFA2/template7.png}
\captionsetup{labelformat=empty}       
 \caption{image-7}
    \end{subfigure}
       \begin{subfigure}[b]{0.24\linewidth}
        \includegraphics[width=\textwidth]{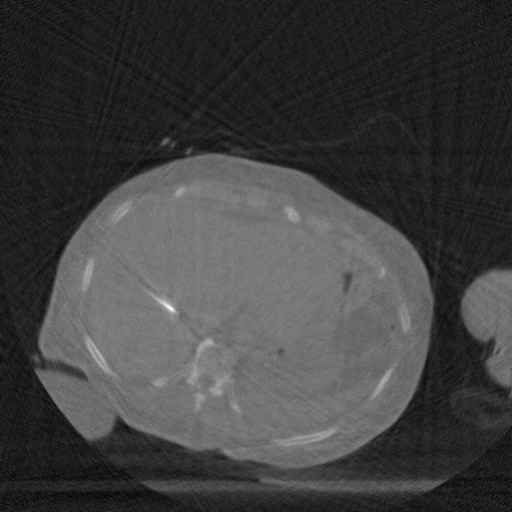}
\captionsetup{labelformat=empty}       
 \caption{image-7 recon}
    \end{subfigure}
 \begin{subfigure}[b]{0.24\linewidth}
        \includegraphics[width=\textwidth]{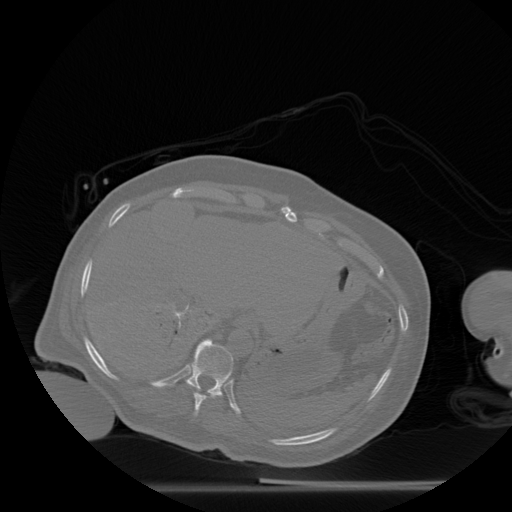}
\captionsetup{labelformat=empty}       
 \caption{image-8}
    \end{subfigure}
       \begin{subfigure}[b]{0.24\linewidth}
        \includegraphics[width=\textwidth]{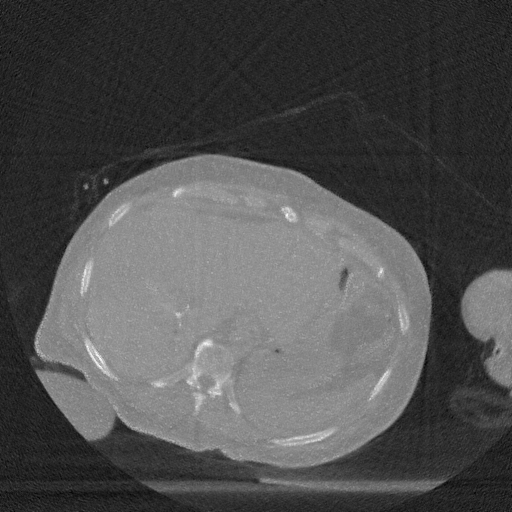}
\captionsetup{labelformat=empty}       
 \caption{image-8 recon}
    \end{subfigure}
       \begin{subfigure}[b]{0.24\linewidth}
        \includegraphics[width=\textwidth]{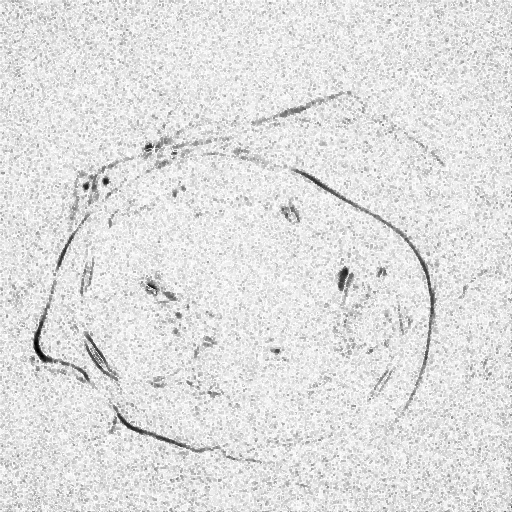}
\captionsetup{labelformat=empty}       
 \caption{weights map}
    \end{subfigure}
       \caption{Reconstructions (referred as `recon') of all slices of  of Fig.~\ref{fig:RFA2_test_templates} using the suggested protocol in Table.~\ref{tab:realistic_imaging_protocol}. The last image is the weights map corresponding to the reconstruction of slice-8. The new changes in the tumor site is picked up by the weights map. }
\label{fig:RFA2_new_protocol}
\end{figure}

Earlier in Sec.~\ref{sec:tmh}, we discussed that the imaging protocols fell under two categories depending on the final goal (tracking or observing details): very few-view imaging and moderate view imaging. \textbf{Ideally, we may prefer to gradually increase from few-view to moderate views as the probe gradually approaches the tumor site.} For the reconstruction of $n^{\textrm{th}}$ slice i.e., slice imaged at time $t=n$, the few-view reconstructions of the previously acquired slices can be used as templates. However, the first scan must always be taken with large number of views because it acts as the initial reference template. Table.~\ref{tab:realistic_imaging_protocol} summarizes this protocol for the dataset of Fig~\ref{fig:RFA2_test_templates}, and Fig.~\ref{fig:RFA2_new_protocol} shows the reconstructions when this realistic protocol is used.

\begin{table}[!h]
\caption{Suggested multi-step imaging protocol for the CT-guided radio-frequency ablation dataset of Fig~\ref{fig:RFA2_test_templates}. The number of views is gradually increased as the probe approaches the tumor site. Only the first scan is taken with large number of views to act as a reference template. The few-view reconstructions of a slice acts as one of the templates for the reconstruction of slice being imaged at next time instant.}
\begin{tabular}{|c|c|c|c|}
\hline
\textbf{\begin{tabular}[c]{@{}c@{}}Slice being\\ imaged at\\ time t\end{tabular}} & \textbf{\begin{tabular}[c]{@{}c@{}}Probe distance from\\ tumor or \\ ablation stage\end{tabular}} & \textbf{\begin{tabular}[c]{@{}c@{}}Number of\\ imaging\\ views\end{tabular}} & \textbf{\begin{tabular}[c]{@{}c@{}}Reconstruction\\ protocol: type \\ of global prior\end{tabular}} \\ \hline
Slice t=1 & Very far from tumor & 360 & CS, no prior \\ \hline
Slice t=2 & Sufficiently far from tumor & 40 & Unweighted \\ \hline
Slice t=3 & Far from tumor & 50 & Unweighted \\ \hline
Slice t=4 & Near tumor & 60 & Unweighted \\ \hline
Slice t=5 & Sufficiently near tumor & 70 & Unweighted \\ \hline
Slice t=6 & Very near tumor & 80 & Unweighted \\ \hline
Slice t=7 & Very near tumor & 90 & Unweighted \\ \hline
Slice t=8 & During, after ablation & 120 & Weighted \\ \hline
\end{tabular}
\label{tab:realistic_imaging_protocol}
\end{table}

\subsection{Effect of hyper-parameters}
\label{subsec:k}
\begin{figure}[h]
    \begin{subfigure}[b]{0.24\linewidth}
        \includegraphics[width=\textwidth]{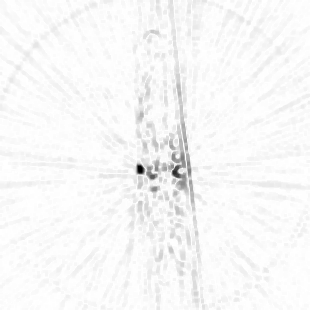}
        \caption{k=10}
     \end{subfigure}     
  \begin{subfigure}[b]{0.24\linewidth}
        \includegraphics[width=\textwidth]{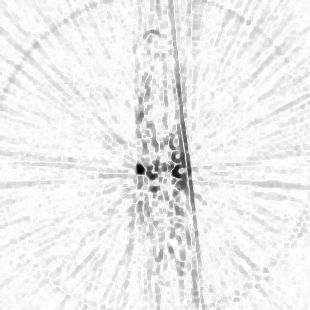}
        \caption{k=50}
     \end{subfigure} 
  \begin{subfigure}[b]{0.24\linewidth}
        \includegraphics[width=\textwidth]{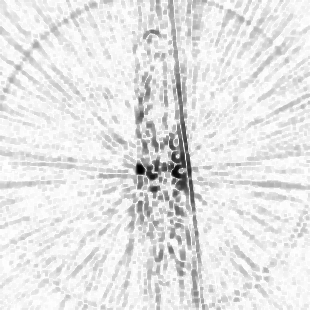}
        \caption{k=90}
     \end{subfigure}
  \begin{subfigure}[b]{0.24\linewidth}
        \includegraphics[width=\textwidth]{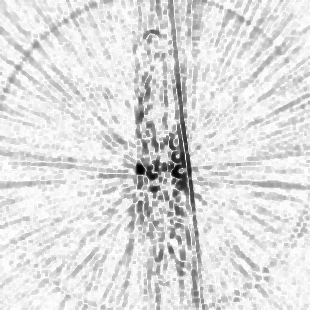}
        \caption{k=130}
     \end{subfigure}
  \begin{subfigure}[b]{0.24\linewidth}
        \includegraphics[width=\textwidth]{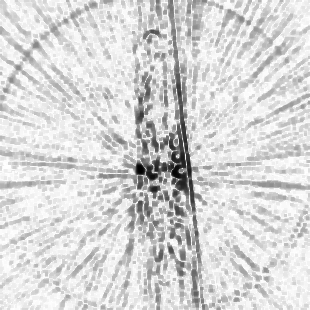}
        \caption{k=170}
     \end{subfigure}
  \begin{subfigure}[b]{0.24\linewidth}
        \includegraphics[width=\textwidth]{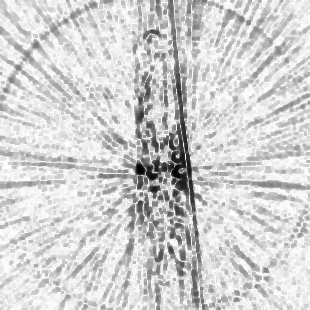}
        \caption{k=210}
     \end{subfigure}
   \begin{subfigure}[b]{0.24\linewidth}
        \includegraphics[width=\textwidth]{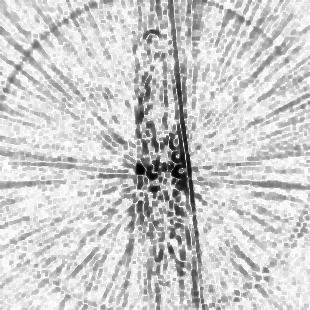}
        \caption{k=250}
     \end{subfigure}
   \begin{subfigure}[b]{0.24\linewidth}
        \includegraphics[width=\textwidth]{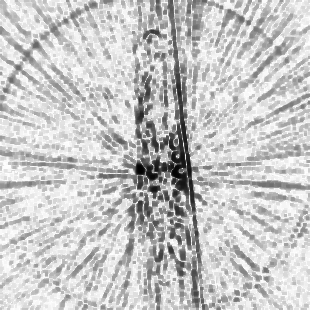}
        \caption{k=290}
     \end{subfigure}
  \begin{subfigure}[b]{0.24\linewidth}
        \includegraphics[width=\textwidth]{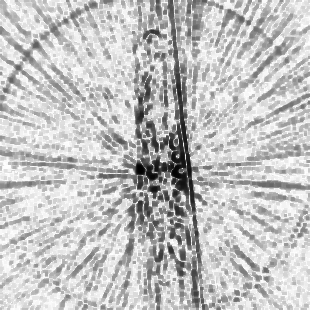}
        \caption{k=330}
     \end{subfigure}
   \begin{subfigure}[b]{0.24\linewidth}
        \includegraphics[width=\textwidth]{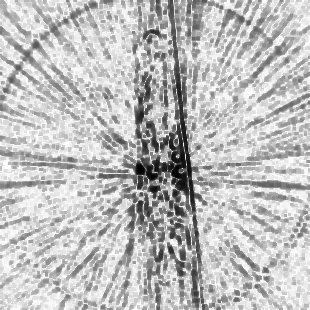}
        \caption{k=370}
     \end{subfigure}
   \begin{subfigure}[b]{0.24\linewidth}
        \includegraphics[width=\textwidth]{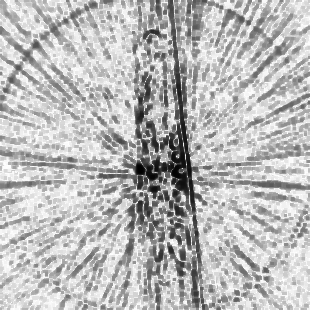}
        \caption{k=410}
     \end{subfigure}
   \begin{subfigure}[b]{0.24\linewidth}
        \includegraphics[width=\textwidth]{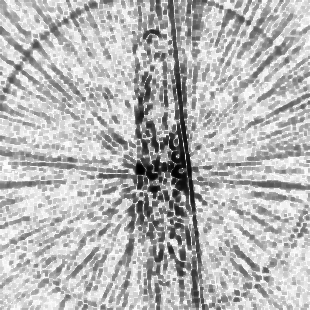}
        \caption{k=450}
     \end{subfigure}
    \caption{Different weight maps for okra reconstruction. Low intensity denotes regions of new changes in test.}
\label{fig:few_view_okra_2D_weights}
\end{figure}

\begin{figure}[h]
    \begin{subfigure}[b]{0.24\linewidth}
        \includegraphics[width=\textwidth]{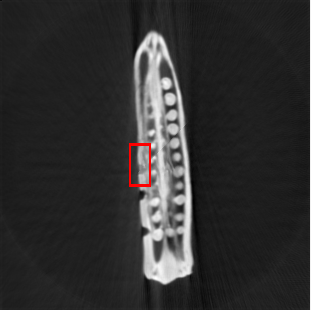}
        \caption{k=10,\\ SSIM = 0.944}
     \end{subfigure}     
  \begin{subfigure}[b]{0.24\linewidth}
        \includegraphics[width=\textwidth]{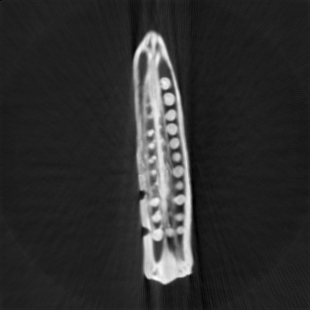}
        \caption{k=50,\\ SSIM = 0.959}
     \end{subfigure} 
  \begin{subfigure}[b]{0.24\linewidth}
        \includegraphics[width=\textwidth]{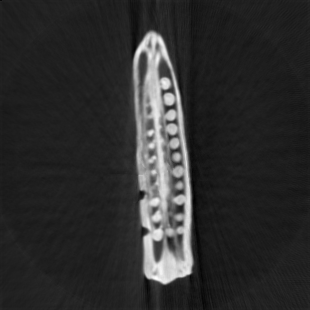}
        \caption{k=90,\\ SSIM = 0.955}
     \end{subfigure}
  \begin{subfigure}[b]{0.24\linewidth}
        \includegraphics[width=\textwidth]{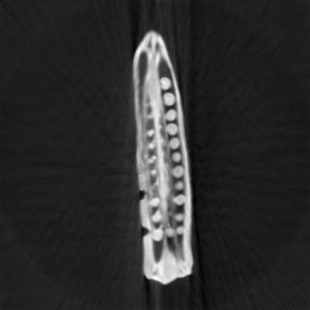}
        \caption{k=130,\\ SSIM = 0.948}
     \end{subfigure}
  \begin{subfigure}[b]{0.24\linewidth}
        \includegraphics[width=\textwidth]{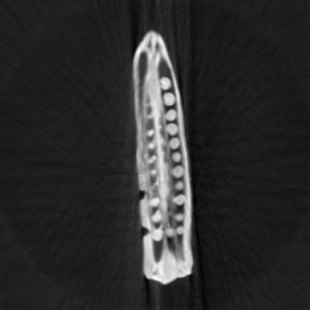}
        \caption{k=170,\\ SSIM = 0.942}
     \end{subfigure}
  \begin{subfigure}[b]{0.24\linewidth}
        \includegraphics[width=\textwidth]{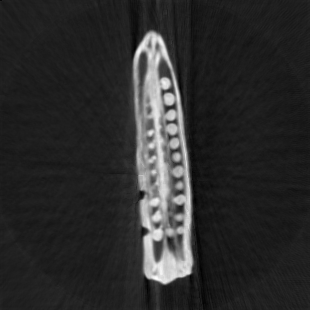}
        \caption{k=210,\\ SSIM = 0.936}
     \end{subfigure}
   \begin{subfigure}[b]{0.24\linewidth}
        \includegraphics[width=\textwidth]{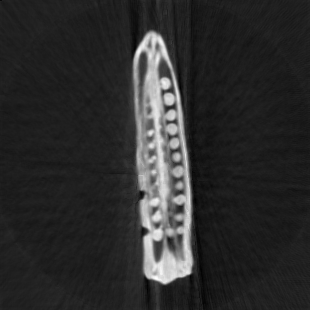}
        \caption{k=250,\\ SSIM = 0.931}
     \end{subfigure}
   \begin{subfigure}[b]{0.24\linewidth}
        \includegraphics[width=\textwidth]{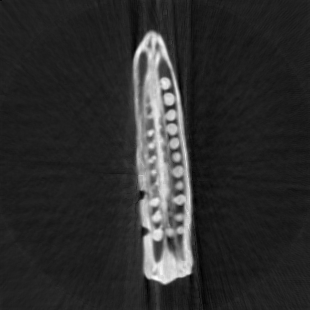}
        \caption{k=290,\\ SSIM = 0.926}
     \end{subfigure}
  \begin{subfigure}[b]{0.24\linewidth}
        \includegraphics[width=\textwidth]{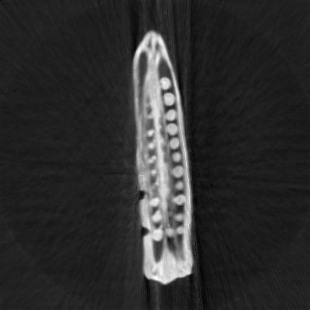}
        \caption{k=330,\\ SSIM = 0.922}
     \end{subfigure}
   \begin{subfigure}[b]{0.24\linewidth}
        \includegraphics[width=\textwidth]{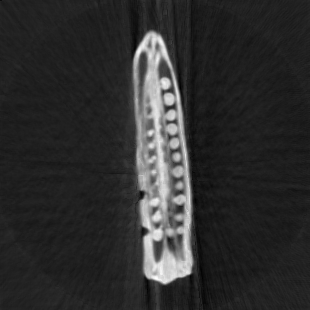}
        \caption{k=370,\\ SSIM = 0.918}
     \end{subfigure}
   \begin{subfigure}[b]{0.24\linewidth}
        \includegraphics[width=\textwidth]{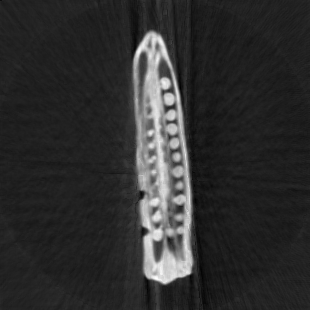}
        \caption{k=410,\\ SSIM = 0.915}
     \end{subfigure}
   \begin{subfigure}[b]{0.24\linewidth}
        \includegraphics[width=\textwidth]{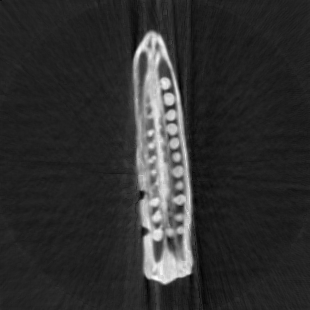}
        \caption{k=450,\\ SSIM = 0.911}
     \end{subfigure}
    \caption{2D reconstructions showing stable reconstructions for large variations in $k$. The SSIM values for all images are computed within the red RoI (shown in (a)), the region where the test is different from all of the templates.}
\label{fig:reconstructions_as_k_varies}
\end{figure}

The parameters $\lambda_1$ and $\lambda_2$ must be chosen empirically or by cross-validation by treating one of the templates as test. In our experiments, we had fixed $\lambda_1$ to be 1 and found that this value is nearly data-independent. The value of $\lambda_2$ largely depends on the amount of artefacts we aim to remove by using prior at the cost of their dominance in the new regions. This value was chosen to lie between $0-1$ for our datasets. Finally, the hyper-parameter $k$ defines the sensitivity of the weights map to the difference between the test image and the prior (projection of test onto the space of templates). When $k=0$, our method converges to the unweighted prior method. As $k$ increases, the weights map starts capturing the new changes in the test, at the cost of detecting a few false positives i.e., false new changes. In other words, as the weights map becomes more sensitive to the difference between the test and templates, it becomes more noisy. In order to visualize the effect of the hyper-parameter $k$, we performed 2D reconstructions of okra dataset for different values of $k$. Fig.~\ref{fig:few_view_okra_2D_weights} shows the weights map obtained for each of the $k$ values. We estimate an approximate choice for the optimal value of $k$ by treating one of the templates as test and reconstructing it. We also note that although the weights map is heavily influenced by $k$, the final reconstructions are stable for large variations in $k$, as seen in Fig.~\ref{fig:reconstructions_as_k_varies}. Alternatively, in cases where one wishes to completely avoid the use of this hyper-parameter, one can construct a binary weights-map using a learning based method described in~\cite{supp_paper}.



\section{Conclusions}
\label{sec:conclusions}
This work maximizes the utility of priors for tomographic reconstruction in longitudinal studies. To the best of our knowledge, no technique in literature addresses the cases when (unweighted) prior-based methods fail. In this work, we demonstrate this and offer a solution. Based on our experiments from radio-ablation clinical data, we show that the choice of either unweighted global prior method presented in~\cite{my_dicta_paper} or the proposed weighted prior method depends on our goal in the application at hand. We establish the context under which these methods can be used, as shown in Schematic~1. When we wish to approximately know the location of new changes, we apply an unweighted prior because it is fast and sufficient for the task at hand. We also choose a smaller number of views in order to reduce radiation. In addition, we show that when our goal shifts to observing the details of the new changes accurately, we acquire projections from a moderate number of views in order to capture more information. We further combine this with the slower but more accurate technique of reconstruction-- the weighted prior-based method. This method ensures that the reconstruction of localized new information in the data is not affected by the priors. We have thus improved the state of the art by detecting these regions of change and assigning low prior weights wherever necessary. The probability of presence of a `new region' is enhanced considerably by a novel combination of different reconstruction techniques.  We have validated our technique on medical 2D and real, biological 3D datasets for longitudinal studies. The method is also largely robust to the number of templates used. We urge the
reader to see the videos of reconstructed volumes in~\cite{supp_paper}. Although the proposed method is built on an eigenspace-based prior, it is infact agnostic to the kind of prior used.

\section{Acknowledgement}
We are grateful to Dr.\ Andrew Kingston for facilitating data collection at the  Australian National University. AR gratefully acknowledges IITB Seed-grant 14IRCCSG012, as well as NVIDIA Corporation for the generous donation of a Titan-X GPU card, which was essential for computing 3D reconstructions for the research reported here. We also thank Dr.\ Akshay Baheti for providing us with medical data from the Tata Memorial Hospital, Mumbai. 

{

\bibliography{cmig_ref}}

\begin{thebibliography}{10}
\expandafter\ifx\csname url\endcsname\relax
  \def\url#1{\texttt{#1}}\fi
\expandafter\ifx\csname urlprefix\endcsname\relax\def\urlprefix{URL }\fi
\expandafter\ifx\csname href\endcsname\relax
  \def\href#1#2{#2} \def\path#1{#1}\fi

\bibitem{barkan17}
O.~{Barkan}, J.~{Weill}, S.~{Dekel}, A.~{Averbuch}, A mathematical model for
  adaptive computed tomography sensing, IEEE Transactions on Computational
  Imaging 3~(4) (2017) 551--565.

\bibitem{fischer16}
A.~Fischer, T.~Lasser, M.~Schrapp, J.~Stephan, P.~B. Noël, Object specific
  trajectory optimization for industrial {X}-ray computed tomography,
  Scientific Reports 6~(19135).

\bibitem{andrei14}
A.~Dabravolski, K.~J. Batenburg, J.~Sijbers, Dynamic angle selection in {X}-ray
  computed tomography, Nuclear Instruments and Methods in Physics Research
  Section B: Beam Interactions with Materials and Atoms 324 (2014) 17 -- 24,
  1st International Conference on Tomography of Materials and Structures.

\bibitem{yang2018}
X.~Yang, V.~De~Andrade, W.~Scullin, E.~L. Dyer, N.~Kasthuri, F.~De~Carlo,
  D.~Gürsoy, Low-dose {X}-ray tomography through a deep convolutional neural
  network, Scientific Reports 8~(1).

\bibitem{geyer2015}
L.~L. Geyer, U.~J. Schoepf, F.~G. Meinel, J.~W. Nance, G.~Bastarrika, J.~A.
  Leipsic, N.~S. Paul, M.~Rengo, A.~Laghi, C.~N. De~Cecco, State of the art:
  Iterative {CT} reconstruction techniques, Radiology 276~(2) (2015) 339--357.

\bibitem{kilic2011}
K.~Kilic, G.~Erbas, M.~Guryildirim, M.~Arac, E.~Ilgit, B.~Coskun, Lowering the
  dose in head {CT} using adaptive statistical iterative reconstruction,
  American Journal of Neuroradiology 32~(9) (2011) 1578--1582.

\bibitem{Shih1992}
S.-C.~B. Lo, S.-L.~A. Lou, S.~K. Mun, Projection domain compensation of missing
  angles for fan-beam {CT} reconstruction, Computerized Medical Imaging and
  Graphics 16~(4) (1992) 259 -- 269.

\bibitem{Donoho}
D.~Donoho, Compressed sensing, IEEE Transactions on Information Theory 52~(4)
  (2006) 1289--1306.

\bibitem{introCS}
E.~Cand{\`e}s, M.~Wakin, An introduction to compressive sampling, IEEE Signal
  Processing Magazine 25~(2) (2008) 21--30.

\bibitem{Varun2013}
V.~P. Gopi, P.~Palanisamy, K.~A. Wahid, P.~Babyn, D.~Cooper, Micro-{CT} image
  reconstruction based on alternating direction augmented lagrangian method and
  total variation, Computerized Medical Imaging and Graphics 37~(7) (2013) 419
  -- 429.

\bibitem{humerus}
{Humerus {CT} dataset}, \url{http://isbweb.org/data/vsj/humeral/}.

\bibitem{PICCS}
G.-H. Chen, J.~Tang, S.~Leng, Prior image constrained compressed sensing
  {(PICCS)}: A method to accurately reconstruct dynamic {CT} images from highly
  undersampled projection data sets, Medical Physics 35~(2) (2008) 660--663.

\bibitem{cardiacPICCS}
G.~Chen, P.~Theriault-Lauzier, J.~Tang, B.~Nett, S.~Leng, J.~Zambelli, Z.~Qi,
  N.~Bevins, A.~Raval, S.~Reeder, H.~Rowley, Time-resolved interventional
  cardiac {C}-arm cone-beam {CT}: An application of the {PICCS} algorithm, IEEE
  Transactions on Medical Imaging 31~(4) (2012) 907--923.

\bibitem{lubner2011}
M.~G. Lubner, P.~J. Pickhardt, J.~Tang, G.-H. Chen, Reduced image noise at
  low-dose multidetector {CT} of the abdomen with prior image constrained
  compressed sensing algorithm, Radiology 260 (2011) 248--256.

\bibitem{pirple}
J.~W. Stayman, H.~Dang, Y.~Ding, J.~H. Siewerdsen, {PIRPLE}: A
  penalized-likelihood framework for incorporation of prior images in {CT}
  reconstruction, Physics in medicine and biology 58~(21) (2013) 7563--82.

\bibitem{mota2017}
J.~F.~C. Mota, N.~Deligiannis, M.~R.~D. Rodrigues, Compressed sensing with
  prior information: Strategies, geometry, and bounds, IEEE Transactions on
  Information Theory 63~(7) (2017) 4472--4496.

\bibitem{Ali2018}
S.~A. Melli, K.~A. Wahid, P.~Babyn, D.~M. Cooper, A.~M. Hasan, A wavelet
  gradient sparsity based algorithm for reconstruction of reduced-view
  tomography datasets obtained with a monochromatic synchrotron-based x-ray
  source, Computerized Medical Imaging and Graphics 69 (2018) 69 -- 81.

\bibitem{liu2016}
J.~Liu, Y.~Hu, J.~Yang, Y.~Chen, H.~Shu, L.~Luo, Q.~Feng, Z.~Gui, G.~Coatrieux,
  3{D} feature constrained reconstruction for low dose {CT} imaging, IEEE
  Transactions on Circuits and Systems for Video Technology PP~(99) (2016)
  1--1.

\bibitem{Xu2012}
Q.~Xu, H.~Yu, X.~Mou, L.~Zhang, J.~Hsieh, G.~Wang, Low-dose {X}-ray {CT}
  reconstruction via dictionary learning, IEEE Transactions on Medical Imaging
  31~(9) (2012) 1682--1697.

\bibitem{my_dicta_paper}
P.~{Gopal}, R.~{Chaudhry}, S.~{Chandran}, I.~{Svalbe}, A.~{Rajwade},
  Tomographic reconstruction using global statistical priors, in: 2017
  International Conference on Digital Image Computing: Techniques and
  Applications (DICTA), 2017, pp. 1--8.

\bibitem{Dong2015}
J.~Dong, W.~Li, Q.~Zeng, S.~Li, X.~Gong, L.~Shen, S.~Mao, A.~Dong, P.~Wu,
  {CT}-guided percutaneous step-by-step radiofrequency ablation for the
  treatment of carcinoma in the caudate lobe, Medicine 94~(39).

\bibitem{tmh}
{Tata Memorial Centre},
  \url{https://en.wikipedia.org/wiki/Tata_Memorial_Centre} (Wikipedia).

\bibitem{rfa}
{Radiofrequency ablation},
  \url{https://en.wikipedia.org/wiki/Radiofrequency_ablation} (Wikipedia).

\bibitem{l1ls}
K.~Koh, S.-J. Kim, S.~Boyd, l1-ls: Simple matlab solver for l1-regularized
  least squares problems, \url{https://stanford.edu/~boyd/l1_ls/} (last
  viewed--July, 2016).

\bibitem{monotone_convergence_theorem}
{Monotone Convergence Theorem},
  \url{https://en.wikipedia.org/wiki/Monotone_convergence_theorem} (Wikipedia).

\bibitem{FDK}
L.~Feldkamp, L.~C. Davis, J.~Kress, Practical cone-beam algorithm, J. Opt. Soc.
  Am 1 (1984) 612--619.

\bibitem{lasso}
R.~Tibshirani, Regression shrinkage and selection via the {Lasso}, Journal of
  the Royal Statistical Society. Series B (Methodological) 58~(1) (1996)
  267--288.

\bibitem{art}
R.~Gordon, R.~Bender, G.~T. Herman, Algebraic reconstruction techniques {(ART)}
  for three-dimensional electron microscopy and {X}-ray photography,
  Theoretical Biology 29~(3) (1970) 471--481.

\bibitem{sart}
A.~Andersen, A.~Kak, Simultaneous algebraic reconstruction technique ({SART}):
  A superior implementation of the {ART} algorithm, Ultrasonic Imaging 6~(1)
  (1984) 81 -- 94.

\bibitem{sirt}
P.~Gilbert, Iterative methods for the three-dimensional reconstruction of an
  object from projections, Journal of Theoretical Biology 36~(1) (1972) 105 --
  117.

\bibitem{supp_paper}
{Supplemental material for `Learning from past scans: Tomographic
  reconstruction to detect new and evolving structures'}, {in Dropbox},
  \url{https://www.dropbox.com/sh/qw5iaukfrkm5o1l/AABt9AE2E7qG57W-uWeS-BPta?dl=0}
  (Including video reconstruction results).

\end{thebibliography}
\end{document}